\documentclass[journal]{IEEEtran}

\usepackage[cmex10]{amsmath}
\usepackage{amsfonts,amssymb,fullpage}
\usepackage{amsrefs}
\usepackage{url} 

\newtheorem{theorem}{\indent Theorem}
\newtheorem{lemma}{\indent Lemma}
\newtheorem{corollary}{\indent Corollary}

\def\Uxoxo{U_{\text{XOXO}}}

\DeclareMathOperator{\tr}{Tr}
\DeclareMathOperator{\sch}{Sch}
\DeclareMathOperator{\CCE}{CCE}
\DeclareMathOperator{\CE}{CE}
\DeclareMathOperator{\clCE}{clCE}
\DeclareMathOperator{\CoCoE}{CoCoE}
\DeclareMathOperator{\clCCE}{clCCE}
\DeclareMathOperator{\clQQE}{clQQE}
\def\geqclean{\stackrel{\!{\scriptstyle clean}}{\geq}\!}
\def\eqclean{\stackrel{\!{\scriptstyle clean}}{=}\!}

\DeclareMathOperator{\poly}{poly}
\DeclareMathOperator{\Var}{Var}
\def\swap{\textsc{swap}}

\def\ba#1\ea{\begin{align}#1\end{align}}
\def\bml#1\eml{\begin{multline}#1\end{multline}}
\def\be{\begin{equation}}
\def\ee{\end{equation}}
\def\bea{\begin{eqnarray}}
\def\eea{\end{eqnarray}}
\def\ben{\begin{eqnarray*}}
\def\een{\end{eqnarray*}}
\def\non{\nonumber}
\def\benum{\begin{enumerate}}
\def\eenum{\end{enumerate}}
\def\bitem{\begin{itemize}}
\def\eitem{\end{itemize}}
\def\l{\left}
\def\r{\right}
\def\>{\rangle}
\def\<{\langle}

\newcommand{\smfrac}[2]{\mbox{$\frac{#1}{#2}$}}
\def\half{\smfrac{1}{2}}
\newcommand{\bra}[1]{\mbox{$\left\langle #1 \right|$}}
\newcommand{\ket}[1]{\mbox{$\left| #1 \right\rangle$}}
\newcommand{\braket}[2]{\mbox{$\langle #1 | #2 \rangle$}}
\newcommand{\dblbraket}[1]{\mbox{$\langle #1 | #1 \rangle$}}
\newcommand{\proj}[1]{\mbox{$| #1 \>\< #1 |$}}
\newcommand{\eq}[1]{Eq.~(\ref{eq:#1})}
\newcommand{\eqs}[2]{Eqns.~(\ref{eq:#1}) and (\ref{eq:#2})}
\newcommand{\sect}[1]{Sec.~\ref{sec:#1}}
\newcommand{\tabref}[1]{Table~\ref{tab:#1}}
\newcommand{\mscite}[1]{Ref.~\cite{#1}}

\newcommand{\thm}[1]{Theorem~\ref{thm:#1}}
\newcommand{\lem}[1]{Lemma~\ref{lem:#1}}
\newcommand{\cor}[1]{Corollary~\ref{cor:#1}}
\newcommand{\upto}[1]{\stackrel{#1}{\approx}}

\newcommand{\myproof}[1]{{\noindent\hspace{2em}{\itshape Proof #1:}}}

\def\ot{\otimes}
\def\eps{{\epsilon}}
\def\cA{{\cal A}}
\def\cB{{\cal B}}

\def\cE{{\cal E}}
\def\cF{{\cal F}}

\def\cN{{\cal N}}
\def\cP{{\cal P}}
\def\bbC{\mathbb{C}}
\def\bbE{\mathbb{E}}
\def\bbZ{\mathbb{Z}}
\def\ra{\rightarrow}
\def\la{\leftarrow}
\def\vb{{\vec{b}}}
\def\vp{{\vec{p}}}
\def\vx{{\vec{x}}}
\def\Cna{{C_1^{(n)}}}
\def\Cnb{{C_2^{(n)}}}

\begin{document}
\title{Time reversal and exchange symmetries of unitary gate capacities}
\author{Aram W.\ Harrow and Peter W.\ Shor%
\thanks{A. W. Harrow is with the Department of Mathematics, University of Bristol, Bristol, BS8 1TW, U.K. 
and is funded by the U.K. EPRSC grant
``QIP IRC'' and the QAP project (contract IST-2005-15848).
email: {\tt a.harrow@bris.ac.uk}}
\thanks{P. W. Shor is with the
Department of Mathematics, Massachusetts Institute of Technology, 
77 Massachusetts Avenue, Cambridge,  
        MA 02139, USA
and is
funded by NSF grant CCF-0431787: ``Quantum Channel Capacities and
Quantum Complexity.''
email: {\tt shor@math.mit.edu}}
}
\date{\today}
\maketitle
\begin{abstract}
Unitary gates are interesting resources for quantum communication in
part because they are always invertible and are intrinsically
bidirectional.  This paper explores these two symmetries:
time-reversal and exchange of Alice and Bob.  We will present examples
of unitary gates that exhibit dramatic separations between forward and
backward capacities (even when the back communication is assisted by
free entanglement) and between entanglement-assisted and unassisted
capacities, among many others.  Along the way, we will give a general
time-reversal rule for relating the capacities of a unitary gate and
its inverse that will explain why previous attempts at finding
asymmetric capacities failed.  Finally, we will see how the ability to
erase quantum information and destroy entanglement can be a valuable
resource for quantum communication.
\end{abstract}

\section{Introduction: Communication using bipartite unitary gates}
\label{sec:intro}
\IEEEPARstart{T}{his} paper investigates the asymptotic communication
capacities of 
bipartite unitary quantum gates; for example, a CNOT
gate with control qubit held by Alice and target qubit held by Bob.
For a review of this topic see \cites{BHLS02,Har05} and references
therein.  The 
question of unitary gate capacity arises when
studying our ability to communicate or generate entanglement using
naturally occurring physical interactions; moreover, studying unitary gate
capacity has often led to new ideas that are useful for other topics in
quantum information theory\cite{Har03}.

In some ways unitary gates are like classical bidirectional channels
or noisy quantum channels, but they are both more complex than one-way
channels because of their intrinsic bidirectionality, and simpler than
noisy channels because they involve no interactions with the
environment.  For example, any nonlocal unitary gate has
nonzero capacities to send classical messages in either
direction and to create entanglement\cites{BHLS02,Beckman01}.  By contrast,
bidirectional classical channels exist that have no capacity in either
direction, but can be useful for nonlocal tasks like reducing
communication complexity\cite{vanDam05}.  Even deterministic
classical bidirectional channels, like the classical CNOT, can
have capacities that are nonzero only in one direction.  Another
feature of unitary gates is that, unlike noisy quantum channels,
knowing the classical capacity (as a function of the amount 
of entanglement assistance) of a unitary gate also determines its
quantum capacity (again parameterized by the amount of entanglement
assistance).  Moreover, allowing free classical communication
does not improve the entanglement capacity; on the other hand, the quantum
capacity appears to no longer be simply equal to the entanglement
generating capacity.  In short, the usual questions (like additivity)
about noisy channel capacities are replaced by an intriguingly
different, yet perhaps related, set of questions about unitary gate
capacities.

In this paper we will investigate the questions of symmetry, both
time-reversal and exchange of Alice and Bob,
that arise in connection with unitary gate capacities.  We will
demonstrate
\begin{itemize}
\item A general rule for relating capacity regions of a gate $U$ to
those of its inverse $U^\dag$ (\sect{reversal}).  Along the way, we
recast the main result of \cite{HL04} as a sort of structure theorem
for communication protocols based on unitary gates, which leads us to
propose a new way to view the capacity region of unitary gates.
\item A gate that exhibits nearly the strongest possible separation
between forward and backward capacities, even when free entanglement
is allowed for back communication (\sect{Vm}).
\item A gate with a nearly maximal separation between its ability to
create and to destroy entanglement.  This gate also exhibits a
near-maximal improvement in communication capacity when assisted by
entanglement (\sect{Vm-dag}).  (A variant of the former result was
independently 
proved for a different gate in \cite{LSW05}.)
\item A quantum communication resource, ``coherent erasure,'' that can
be thought of as the time-reversal of coherent classical
communication (\sect{erasure}).
\item A more restricted type of resource inequality, which we call a
  ``clean resource inequality.'' (described in more detail below)
\item Alternate proofs for the two main unitary gate capacity theorems
that are currently known (the Appendix
).  The proofs are
simpler and establish slightly stronger versions of the capacity
theorems; we also include them because they make this paper a
self-contained summary of almost every result to date on asymptotic
unitary gate capacities.
\end{itemize}
As we will see, a number of speculative claims about unitary gate capacities
remain to be fully resolved, and more interestingly, we are only
beginning to pose our questions about them in the right way.
We conclude in \sect{conclude} with some ideas about future research.

The remainder of this section reviews notation and some background
results.  Following \cites{DHW05,Har05} we state our coding theorems in
the language of asymptotic resource inequalities.  The basic
asymptotic resources are $[c\ra c]$ (one use of a noiseless classical
channel from Alice to Bob, a.k.a. a cbit), $[qq]$ (the state
$\ket{\Phi}^{AB} = \frac{1}{\sqrt{2}}\sum_{x=0}^1 \ket{x}^A\ket{x}^B$,
a.k.a. an ebit) and $[q\ra q]$ (one use of a noiseless quantum
channel, a.k.a. a qubit).  Protocols transforming these resources into
one another (e.g. teleportation) are expressed as {\em
asymptotic resource inequalities} such as $2[c\ra c]+[qq]\geq [q\ra q]$.
We will also make use of coherent
bits, or cobits, which are denoted $[q\ra qq]$ and
correspond to the isometry $\sum_{x=0}^1
\ket{x}^A\ket{x}^B\bra{x}^A$.  Coherent bits were introduced in
\cite{Har03} which proved that $2[q\ra qq]=[q\ra q]+[qq]$ (though only
as an asymptotic relation; see \cite{vanEnk05} for a single-shot version).
Since we are interested in two-way
communication, define $[c\la c]$, $[q\la q]$ and $[qq\la q]$ to be
cbits, qubits, and cobits, respectively, sent from Bob to Alice.
These definitions are summarized in \sect{notation}.

For a unitary gate $U$, let $\<U\>$ denote the corresponding
asymptotic resource; we can use it to state resource inequalities such
as $\<CNOT\>\geq [c\ra c]$.  Define $\CCE(U)$ to be the
three-dimensional capacity region of $U$ to send cbits forward, send
cbits backwards and generate entanglement:
\ba \CCE(U) &:= \{ (C_1,C_2,E) : \non\\
&\<U\> \geq C_1 [c\ra c] + C_2[c\la c]
+ E[qq]\}.\ea
For example, if we define $\swap$ to exchange a qubit of Alice's with
a qubit of Bob's, then $(1,1,0)\in\CCE(\swap)$, since one use of \swap can be
used to send one bit forward and another bit backwards at the same time.
When $C_1,C_2,E$ take on negative values, we move the corresponding
resources to the left-hand side of the resource inequality, e.g. $E<0$
for entanglement-assisted communication.  Continuing with the \swap
example, we could use super-dense coding to consume 2 ebits and send 2
cbits in either direction: thus $(2,2,-2)\in\CCE(\swap)$.
We can define capacities in
terms of $\CCE(U)$ as extremal points of the region: the entanglement
capacity $E(U):=\max\{E : (0,0,E)\in\CCE(U)\}$, the forward classical
capacity $C_\ra(U):=\max\{C : (C,0,0)\in \CCE(U)\}$, the backwards
capacity $C_\la(U):=\max\{C : (0,C,0)\in \CCE(U)\}$, the simultaneous
capacity $C_+(U):=\max\{C_1 + C_2 : (C_1,C_2,0)\in \CCE(U)\}$, and
entanglement-assisted versions $C_\ra^E(U):=\max\{C : (C,0,-\infty)\in
\CCE(U)\}$, $C_\la^E(U):=\max\{C : (0,C,-\infty)\in \CCE(U)\}$ and
$C_+^E(U):=\max\{C_1 + C_2 : (C_1,C_2,\infty)\in \CCE(U)\}$.  In
\cites{BHLS02,Beckman01}, it was shown that one of these capacities is
nonzero if 
and only if all of them are nonzero.  Various quantitative relations
among these 
capacities were also shown, but they will be subsumed in what follows.

We can analogously define the capacity region of achievable rates of
entanglement generation and {\em coherent} communication in both
directions.  In \cite{HL04} this region was shown to coincide with
$\CCE(U)$ for the $C_1,C_2\geq 0$ quadrant, and to be trivially
related for other quadrants.  Here we will present this result in a
slightly stronger form.  If $\alpha=(\alpha_n)_{n=1}^\infty$ and
$\beta=(\beta_n)_{n=1}^\infty$ are two {\em pure}\footnote{Following
  \cite{DHW05} we say that a resource is pure if it is an isometry or
  a pure state.} asymptotic resources such that $\alpha\geq \beta$,
then we say that this resource inequality is clean (denoted
$\alpha\geqclean\beta$) if $\alpha_n$ can be mapped to
$\beta_{n(1-\delta_n)}$ using a protocol that discards only
$n\delta_n'$ qubits, which (up to error $\epsilon_n$) are all in the
state $\ket{0}$, where $\epsilon_n,\delta_n,\delta_n'\ra 0$ as
$n\ra\infty$.\footnote{The idea that discarding information should be
  costly dates back to Szilard's interpretation in 1929 of Maxwell's
  demon\cite{Szi29} (see also \cite{Bennett82}), and in the context of
  quantum Shannon theory has been discussed in \cite{OHHH01}.}
 We will also call protocols ``semi-clean'' when, at the end
of the protocol, they discard arbitrary $n\delta_n'$-qubit states
which depend only on $n$ and not on any other inputs or outputs of the
protocol.  In most cases of interest, semi-clean protocols are also clean.

  Implicit in the definition of a clean protocol
is the idea that all
local quantum operations are represented by isometries (perhaps
increasing the dimension) followed by discarding some qubits.
The advantage of this
formulation is that apart from the discarding step, protocols can be
easily reversed. 
On the other hand, requiring that resources be pure is quite a
restrictive condition, and 
hopefully future work will able to fruitfully relax it.  Many
common resource inequalities, such as entanglement
concentration/dilution, remote state preparation, channel coding,
etc..., can be shown to admit ``clean'' versions, but other simple
inequalities such as $2[qq]\geq [qq]$ do not.  
Now define
\bml\clCCE(U) := \{(C_1,C_2,E) : \\
\<U\> \geqclean C_1 [q\ra qq]
+ C_2 [qq\la q] + E[qq]\}.\eml
Refs \cites{HL04,Har05} considered the
similar region $\CoCoE(U)$ in which there was no requirement that the
protocols be clean.  Although $\clCCE(U)$ is still convex, it is no
longer monotone in the sense that throwing away resources does not
always yield 
 valid protocols; for example, while points like $(0,0,-\infty)$
are in $\CCE(U)$, one can show that $E(U^\dag) = \max\{E :
(0,0,-E)\in\clCCE(U)\}$.  This result can be proven directly using the
formula for $E(U)$ in \cites{BHLS02,Leifer03}, and
will also follow from a more
general theorem relating $\clCCE(U)$ to $\clCCE(U^\dag)$ that we prove
in \sect{reversal}.

The main use of $\clCCE$ in this paper will be the following
strengthening of \cite{HL04}'s main result.
\begin{theorem}[std. form of unitary protocols]
\label{thm:std-form}
If 
$$\<U\> \geq C_1 [c\ra c] + C_2 [c\la c] + E[qq]$$
 and $C_1,C_2\geq 0$, then there exists $E'\geq E$ such that
$$\<U\> \geqclean C_1 [q\ra qq] + C_2 [qq\la q] + E'[qq].$$
  In other words,
there exists a series of protocols $(\cP_n)_{n=1}^{\infty}$, with
\be\cP_n = (A_n^{(n)}\ot B_n^{(n)}) U 
\ldots U(A_1^{(n)}\ot B_1^{(n)}) U (A_0^{(n)}\ot B_0^{(n)})
\label{eq:std-form1}\ee
 for local
isometries $A_0^{(n)},B_0^{(n)},\ldots,A_n^{(n)},B_n^{(n)}$
(possibly adding ancillas), such that for all 
$x\in\{0,1\}^{n(C_1-\delta_n)}, y\in\{0,1\}^{n(C_2-\delta_n)}$, we have
\bml
\cP_n \ket{x}^A \ket{y}^B \upto{\epsilon_n}\\
\ket{x,y}^A \ket{x,y}^B (\ket{\Phi}^{AB})^{\ot n(E-\delta_n)}
(\ket{00}^{AB})^{\ot ^{n\delta'_n}}
.\label{eq:std-form2}\eml
Here
$\epsilon_n,\delta_n,\delta_n'\ra 0$ as $n\ra \infty$
and $\rho\upto{\eps}\sigma$ means that 
$\half\|\rho-\sigma\|_1\leq \epsilon$.

In terms of $\CCE$ and $\clCCE$, this means that for any
$(C_1,C_2,E)\in \CCE$ there exists $E'\geq E$ such that 
$(C_1,C_2,E')\in \clCCE$.
\end{theorem}
\mscite{Har05} sketched how to extend the proof of \cite{HL04} to
obtain the above theorem, but we will make use of \cite{Har08a} to
give a more rigorous derivation in the Appendix.

Finally, we state a single-shot expression for the trade-off curve
between ebits and cbits sent from Alice to Bob\cites{Har03,Har05}; call
this tradeoff curve $\CE(U)$ and define it to be  $\CE(U) := \{(C,E) :
(C,0,E)\in\CCE(U)\}$.  Similarly define $\clCE(U) :=
\{(C,E) : (C,0,E)\in\clCCE(U)\}$.
Before we can state our expression for
$\clCE(U)$, we will need a few more definitions (following
\cite{DHW05}).  
For a state $\psi^{ABC}=\proj{\psi}^{ABC}$, recall the definition of
the von Neumann 
entropy as $H(A) = H(A)_\psi = H(\psi^A) = -\tr (\psi^A \log \psi^A)$,
where $\psi^A = \tr_{\!BC} \,{\psi}^{ABC}$.  Similarly the quantum
mutual information is $I(A;B)=H(A)+H(B)-H(AB)$ and the conditional
information is defined as $H(B|A) = H(AB) - H(A)$.  Here, and
elsewhere, we omit subscripts when the underlying state is obvious.
We will denote an ensemble of pure states by $\cE^{XABA'B'} = \sum_x
p_x \proj{x}^X \ot \proj{\psi_x}^{ABA'B'}$.  Here $X$
is a classical label, the gate $U$ acts on $AB$, and $A'B'$ are
ancilla systems of arbitrary finite dimension.  Let $U(\cE)$ stand for
$(U^{AB} \ot I^{XA'B'})(\cE)$.  In terms of these ensembles we can define
\bml \Delta_{I,E}(U) := \l\{ (C,E) : \exists \cE \mbox{ s.t. }\right.\\
I(X ; BB')_{U(\cE)} - I(X ; BB')_{\cE} \geq C \\
\left. \text{ and }H(BB'|X)_{U(\cE)} - H(BB'|X)_{\cE} = E\r\},\eml
where $\cE$ is an ensemble of bipartite pure states in $ABA'B'$ 
conditioned on a classical register $X$.  This corresponds to the
set of single-shot increases in mutual information ($I(X;BB')$) and
average entanglement ($H(BB'|X)$) that are possible.  It turns out
that these increases are also achievable asymptotically, as expressed
in the following theorem:
\begin{theorem}\label{thm:C-E-toff}
 $\clCE(U)$ is equal to the closure of $\Delta_{I,E}(U)$.
\end{theorem}
The direct coding theorem was proven in \cite{Har03} and the converse
in \cite{Har05}*{Section 3.4.2}.  In the Appendix 
we will give a new, and more self-contained, proof of the coding
theorem.

\section{Reversing unitary communication protocols}\label{sec:reversal}
In this section, we present a general theorem for relating the
capacity region of $U$ with the capacity region of $U^\dag$.
Many of the key ideas are illustrated by the gate
$\Uxoxo$, which was conjectured in \cite{BHLS02} to have asymmetric
communication capacities.  $\Uxoxo$ acts on a $2^m\times 2^m$-dimensional
space, with $m$ a parameter, and is defined as
\ben
\Uxoxo \ket{x0} &=& \ket{xx} \qquad\forall x\in\{0,1\}^m\\
\Uxoxo \ket{xx} &=& \ket{x0} \qquad\forall x\in\{0,1\}^m\\
\Uxoxo \ket{xy} &=& \ket{xy} \qquad\forall x\neq y \in\{0,1\}^m
\een
Clearly $\<\Uxoxo\>\geq m[q\ra qq]$, but at first glance it appears
that $\Uxoxo$ cannot easily be used for communication from Bob to
Alice. Indeed \cite{Har05} proved that when starting without
correlation or entanglement, a single use of $\Uxoxo$ could not send
more than $0.7m + o(m)$ bits from Bob to Alice.

However, by consuming entanglement, $\Uxoxo$ can be used to send $m$
bits from Bob to Alice.  The protocol is as follows:
\benum\item Start with
$2^{-m/2}\sum_{x\in\bbZ_2^m}\ket{\vx}^A\ket{\vx}^B$.
\item
To encode
message $b_1,\ldots,b_m$, Bob applies $Z_1^{b_1}\ldots Z_m^{b_m}$ and
obtains the state $2^{-m/2}\sum_{x\in\bbZ_2^m} (-1)^{\vb\cdot \vx}
\ket{\vx}^A\ket{\vx}^B$
\item $\Uxoxo$ is applied to yield the state
$2^{-m/2}\sum_{x\in\bbZ_2^m} (-1)^{\vb\cdot \vx}\ket{\vx}^A\ket{0}^B$.
\item Alice applies $H^{\ot m}$ and obtains $\ket{\vb}^A\ket{0}^B$.
\eenum
Thus $\<\Uxoxo\> + m[qq]\geq m[qq\la q]$.  As a corollary,
$C_\la(\Uxoxo)\geq m/2$.  If we could prove that this were roughly
tight (say that $C_\la(\Uxoxo)\leq m/2 + o(m)$), then we might
conclude that
forward and backward capacities can be separated by a constant factor,
but that this separation vanishes when entanglement is allowed for
free.  We will later demonstrate unitary gates with much stronger
separations, even between entanglement-assisted capacities.

First, we can generalize the backward communication protocol of
$\Uxoxo$ to obtain the following result:
\begin{theorem}\label{thm:reversal}
$(C_1,C_2,E)\in\clCCE(U)$ $\;\Leftrightarrow\;$
$(C_2,C_1,-E-C_1-C_2)\in\clCCE(U^\dag)$ 
\end{theorem}
\begin{proof}
The proof follows almost immediately from \thm{std-form}.  Suppose
$(C_1,C_2,E)\in\clCCE(U)$, so that for any $\delta,\eps>0$ and all
$n$ sufficiently large there exists a protocol $\cP_n$ of the form
of \eq{std-form1} that satisfies \eq{std-form2}. 
We can assume WLOG that the local isometries in \eq{std-form1} are in fact
unitaries, with all the ancillas being added at the beginning. 
Then we take the complex
conjugate of \eq{std-form1} to obtain
\be\cP_n^\dag = 
(A_0^{(n)}\ot B_0^{(n)})^\dag U^\dag 
 \dots
U^\dag (A_n^{(n)}\ot B_n^{(n)})^\dag.
\ee
Observe that  $\cP_n^\dag$ is a protocol that uses $U^\dag$ $n$
times together with with local resources.  Alice and Bob have only to
create the $n\delta_n'$ copies of 
$\ket{00}$, which they can do using local isometries for free.  Then
they can use
$U^\dag$ $n$ times to apply $\cP_n^\dag$. Up to error $\eps_n$, this maps
$\ket{x}^A\ket{x}^B\ra \ket{x}^A$ for $x\in\{0,1\}^{n(C_1-\delta_n)}$ and
$\ket{y}^A\ket{y}^B\ra \ket{y}^B$ for $y\in\{0,1\}^{n(C_2-\delta_n)}$,
while consuming $n(E+\delta_n)$ ebits (or generating $-n(E+\delta_n)$ ebits).
By applying the protocol outlined above for $\Uxoxo$, this can be used
to send $n(C_2-\delta_n)$ cobits from Alice to Bob and
$n(C_1-\delta_n)$ cobits from Bob to Alice, while consuming
$n(C_1+C_2+E-\delta_n)$ ebits (or generating $-n(C_1+C_2+E-\delta_n)$
ebits).
\end{proof}

From the definition of entanglement capacity we now obtain a statement
claimed in the last section.
\begin{corollary}\label{cor:ent-dest-cap}
For any unitary $U$, $E(U^\dag) = \max\{E : (0,0,-E)\in\clCCE(U)\}$.
\end{corollary}

We can also obtain a few immediate corollaries for the case of free
entanglement.
\begin{corollary}\label{cor:sym-cap}
For any unitary $U$, 
\begin{enumerate}
\item $(C_1,C_2,-\infty)\in\CCE(U) \iff
(C_2,C_1,-\infty)\in\CCE(U^\dag)$.
\item In particular, $C_\ra^E(U) = C_\la^E(U^\dag)$
\item If $U=U^\dag$ then $C_\la^E(U)=C_\ra^E(U)$.
\item If $U=U^\dag$ then $C_\ra(U) \geq C_\ra^E(U)/2$.
\end{enumerate}
\end{corollary}

The only nontrivial claim here is (4).  To prove it, first note that
the entanglement assisted capacity can be achieved with the assistance
of $\leq E(U^\dag)$ ebits.  This is because $(C,0,-E)\in\clCCE(U)$ implies
$(0,0,-E)\in\clCCE(U)$ (since cobits can always be discarded cleanly)
and \thm{reversal} implies that $E(U^\dag)\geq E$.  Since we have
assumed $U=U^\dag$, we have $E(U)=E(U^\dag)\geq E$.  Thus two uses of
$U$ can 
send $C$ cobits: the first use generates $E$ ebits and the second use
consumes them to send $C$ cobits.  (A similar result was proved in
\cite{BS03b}.)

Note that cases (3) and (4) of \cor{sym-cap} apply to $\Uxoxo$, and
show us why we should not expect a dramatic separation of capacity for
$\Uxoxo$, or indeed any gate equal to its inverse. However, a
straightforward modification of the argument in the next section can
be used to prove \cite{BHLS02}'s conjecture that $C_\la(\Uxoxo)\leq
m/2 + o(m)$ for $m$ large, which nearly saturates the bound
$C_\la(U)\geq C_\ra(U)/2$ that we now understand for gates satisfying
$U=U^\dag$.

An alternate proof of the reversal theorem can be obtained from the resource
equality $2[q\ra qq]\eqclean [q\ra q] + [qq]$.  \thm{reversal} is then
equivalent to the claim that $(Q_1,Q_2,E)\in\clQQE(U) \Leftrightarrow
(Q_2,Q_1,-E)\in\clQQE(U^\dag)$.  See \cite{Devetak06} for similar
examples of reversing quantum communication protocols.

\section{$V_m$: A gate with asymmetric capacities}\label{sec:Vm}
Guided by \thm{reversal}, we will construct a gate that is quite
different from its inverse.  Again choosing a positive integer $m$ as
a parameter, define $V_m$ on $\bbC^{2^m}\ot \bbC^{2^m}$ by
\ben
V_m \ket{x}^A\ket{0}^B &=& \ket{x}^A\ket{x}^B
 \qquad\qquad\forall x\\
V_m \ket{x}^A\ket{y}^B &=& \ket{x}^A\ket{y-1}^B 
\qquad \mbox{for } 0<y\leq x\\
V_m \ket{x}^A\ket{y}^B &=& \ket{x}^A\ket{y}^B 
\qquad\qquad \mbox{for } y>x
\een
The first line means that $\<V_m\>\geq m[q\ra qq]$.  Though we will
not need this fact, it turns out that $C_\ra(V_m)=m$.  The proof,
following a similar argument for $\Uxoxo$ in \cite{BHLS02} is as
follows.  Since $V_m = \sum_x \proj{x} \ot B_x$ for some operators
$B_x$, we know that the
Schmidt rank of $V_m$ is $\leq 2^m$.  Thus
$$m\leq C_\ra(V_m) \leq E(V_m) \leq \log \sch(V_m)\leq m$$
and each inequality must be an equality.

To bound the back communication of $V_m$, we now claim that $V_m$ can
be simulated to within an accuracy of $\eps$ by using $m[q\ra qq]$ and
$O(\log m/\eps) ([q\ra q]+[q\la q])$.  We would like
to say that this implies
\bml m[q\ra qq] + O(\log m)([q\ra q]+[q\la q]) 
\\ \geq \<V_m\> \geq 
m[q\ra qq].\label{eq:desired-Vm-bounds},\eml
but our tools are not strong enough to actually prove this.  This is
because simulating $n$ copies of $V_m$ to constant accuracy would
require $O(n(m+\log nm))$ qubits of communication, which is
superlinear in $n$ for any fixed $m$.  However, for now suppose that
\eq{desired-Vm-bounds} were true.  It would imply that
\be C_\la^E(V_m)\leq O(\log m) \ll m = C_\ra(V_m),
\label{eq:Vm-sep}\ee
 a rather dramatic separation between forward and backward capacities,
even when we allow free entanglement to assist the back communication.
By using techniques specialized to unitary gates, we will give a proof of
\eq{Vm-sep} later in this section; the proof is inspired by
\eq{desired-Vm-bounds}, but of course does not rely on it.

Our simulation also means that $V_m$ could be
thought of as almost equivalent, at least for large $m$, to the
resource of
coherent classical communication. This is interesting both because it
is more natural to implement cobits as a unitary gate than as an
isometry and because unitary gates, unlike isometries, are reversible.
We will return to this second point in the next section when we
discuss $V_m^\dag$.  However, we cannot state this as a more precise
statement about asymptotic resources since the sequence
$(V_m)_{m=1}^\infty$ does not fit \cite{DHW05}'s definition of an
asymptotic resource.

\subsection{A simulation for $V_m$}
In this section, we show how $V_m$ can be simulated up to error $\eps$
by a protocol that uses $m[q\ra qq]$ and $O(\log m/\eps) ([q\ra
q]+[q\la q])$.  A key subroutine used in the simulation is a classical
communication protocol for distributed comparison.  Suppose
$x,y\in\{0,1\}^m$, Alice holds $x$ and Bob holds $y$, which we
interpret as integers between 0 and $2^m-1$.  Then for any error
probability $\eps>0$ they can probabilistically determine whether
$x=y$, $x>y$ or $x<y$ using $O(\log m/\eps)$ bits of
communication~\cite{Nisan93}.  The comparison protocol is designed for
classical information, but our simulation of $V_m$ will run it
coherently using quantum communication.

For the simulation for $V_m$,  suppose Alice and Bob start with
$\ket{x}^{A_1}\ket{y}^{B_1}$ for 
$x,y\in\{0,1\}^m$.  Our protocol is as follows.
\benum
\item  Define the indicator variable $w$ to be 1 if $y=0$, 2 if $0<y\leq
x$ or 3 if $y>x$. 
Use $O(\log m/\eps)$ qubits of communication
in either direction to coherently compute $w$.  
This leaves them (up
to error $\eps$) with the state
$$\ket{x}^{A_1}\ket{y}^{B_1}\ket{w}^{A_2}\ket{w}^{B_2}
\ket{f(x,y)}^{A_3B_3},$$
 and where $f(x,y)$ is the state of the ancillas
produced by the comparison subprotocol. 
Using a standard
procedure (compute $\ket{w}$, copy it to a new register and then
uncompute the first copy of $\ket{w}$ along with the ancilla states
produced along the way), Alice and Bob can eliminate the ancilla
register to hold simply
$$\ket{x}^{A_1}\ket{y}^{B_1}\ket{w}^{A_2}\ket{w}^{B_2},$$
again up to error $\eps$.
\item Use $m [q\ra qq]$ as follows:
\bitem
\item If $w=1$, then Alice inputs $\ket{x}^{A_1}$, which maps to the state
$\ket{x}^{A_1}\ket{x}^{B_3}$.  Since $y=0$, the $B_1$ register is in
the $\ket{0}$ state.  Bob swaps $B_1$ and $B_3$ to obtain
$\ket{x}^{B_1}\ket{0}^{B_3}$.
\item If $w=2$ or 3, then Alice inputs $\ket{0}^{A_3}$, which maps to the state
$\ket{0}^{A_3}\ket{0}^{B_3}$.  Then she discards $A_3$.
\eitem
In either case Bob discards register $B_3$, which always contains the
$\ket{0}$ state.
\item If $w=2$ then Bob maps $\ket{y}^{B_1}$ to $\ket{y-1}^{B_1}$.
\item Alice and Bob use an additional $O(\log m/\eps)$ qubits of communication to uncompute
$\ket{w}^{A_2}\ket{w}^{B_2}$.
This point is slightly subtle, as the meaning of $w$ as changed: now
$w=1$ means that $y=x$, $w=2$ means that $y<x$ and $w=3$ means that
$y>x$.  Thus, Alice and Bob will compute
$\ket{w'}^{A_3}\ket{w'}^{B_3}$ corresponding to these new cases, and
will each map $\ket{w}\ket{w'}$ to $\ket{w\oplus w'\pmod{3}}\ket{w'}$.
With probability $\geq 1-2\eps$ we have $w=w'$, so this operation
effectively erases $\ket{w}$. Then they uncompute
$\ket{w'}^{A_3}\ket{w'}^{B_3}$, along with all the ancilla produced
along the way.
\eenum

The entire procedure uses $m[q\!\ra\! qq]\! +\! O(\log m/\eps)([q\!\ra\! q] \!+\! [q\!\la\! q])$. 
To see that the protocol works, first observe that in an ideal
protocol where equality testing was perfectly accurate we would obtain
precisely $V_m$.  Thus when we replace equality testing with an
approximate version that is accurate (in the sense of
cb-norm\cite{KW03}) to within $\eps$, the overall protocol has error
$\leq 2\eps$.

\subsection{Bounding the backwards capacity of $V_m$}
We now use our simulation for $V_m$ to prove that $C_\la^E(V_m)\leq
O(\log m)$.  Since the cost of our simulation depends on the
desired error rate, standard resource arguments are not enough to
prove this claim.  Instead, let $V_m^\eps$ denote the result of
simulating $V_m$ to accuracy $\eps$ using the above procedure.  Since
$V_m^\eps$ is built out of cobits and qubits, it will be an isometry
rather than a unitary operator.  This is not a serious problem:
\thm{C-E-toff} still applies and by extending the space that $V_m$
acts on, we have 
$\|V_m - V_m^\eps\|_\infty \leq \eps$, where $\|X\|_\infty :=
\max_{\ket{\psi}: \dblbraket{\psi}=1} \sqrt{\bra{\psi}X^\dag
  X\ket{\psi}}$.  
Moreover, we can
simulate $V_m^\eps$ 
{\em exactly} using $m[q\ra qq] + O(\log m/\eps)([q\ra
  q]+[q\la q])$.  Thus $C_\la^E(V_m^\eps) \leq O(\log m/\eps)$.  Then
  we will conclude that $C_\la^E(V_m) \leq 
O(\log m)$ by choosing $\eps = 1/m$ and applying the following
lemma:

\begin{lemma}[Continuity of one-way capacity]\label{lem:continuity}
If $U$ and $V$ are isometries with outputs in $\bbC^d\ot \bbC^d$ such
that $\|U-V\|_{\text{cb}} \leq \eps$ then for all $(C,E)\in\clCE(U)$
there 
exists $(C',E')\in\clCE(V)$ such that $|C-C'|\leq\eps'$ and $|E-E'|\leq
\eps'$, where $\eps'=8\eps\log d + 4H_2(\eps)$ and 
$H_2(\eps):=-\eps\log \eps - (1-\eps)\log(1-\eps)$.
\end{lemma}
\begin{proof}
By \thm{C-E-toff}, for any $\delta>0$ there exists an ensemble of
bipartite 
pure states $\cE^{XABA'B'}$ such that
\ba I(X;BB')_{U(\cE)} - I(X;BB')_{\cE} &\geq C-\delta
\\
\l|H(BB'|X)_{U(\cE)} - H(BB'|X)_{\cE} - E \r| &\leq \delta,
\ea
where $U$ acts on the $d$-dimensional systems $A$ and $B$, while $X,
A'$ and $B'$ can have arbitrarily large dimension.  Thus we will need to use
a recently proved variant of Fannes' inequality\cite{AF04} which
bounds the change in relative entropy as a function of the dimension
of only the first system.   Begin by using the chain rule\cite{CT91}
to express $H(BB'|X)$ as
$H(B|B'X) + H(B'|X)$.  Ref.~\cite{AF04} states that if
$\|U(\cE)-V(\cE)\|_1 \leq \eps$ then
$$\l|H(B|B'X)_{U(\cE)} - H(B|B'X)_{V(\cE)} \r| \leq 2H_2(\eps) +
4\eps\log d,$$
where $d=\dim B$.
On the other hand $\cE^{B'X} = U(\cE)^{B'X} = V(\cE)^{B'X}$, so
$H(B'|X)_{U(\cE)} = H(B'|X)_{V(\cE)}$ and 
$$ \l|H(BB'|X)_{U(\cE)} - H(BB'|X)_{V(\cE)}\r| \leq 2H_2(\eps) + 4\eps
\log d.$$
Similarly $I(X;BB') = H(BB') - H(BB'|X) = H(B') + H(B|B') - H(BB'|X)$ and $H(B')$ is unchanged by
applying a unitary to $AB$, so 
$$ \l|I(X;BB')_{U(\cE)} - I(X;BB')_{V(\cE)}\r| \leq 4H_2(\eps) + 8\eps
\log d.$$
If we now take $\delta\ra 0$
and apply \thm{C-E-toff} again to
relate $\Delta_{I,E}(V)$ to $\clCE(V)$, we obtain the
proof of the lemma.
\end{proof}

{\em Remark:}  We suspect that the entire two-way communication
capacity region $\CCE(U)$ is similarly continuous.  However,
without a characterization of the two-capacity analogous to
\thm{C-E-toff}, the proof technique used in Lemma~\ref{lem:continuity}
will not work.

\section{More asymmetry: the capacity region of $V_m^\dag$}
\label{sec:Vm-dag}
In this section we will demonstrate 
nearly tight bounds for the capacity region of $V_m^\dag$, just as we
did with $V_m$.  We will find large  separations between
entanglement-assisted and -unassisted capacities, as well as between
entanglement-creation and -destruction capacities.

First use \thm{reversal} to show that $\<V_m^\dag\> + m[qq] \geq
m[qq\la q]$, or equivalently, that $\<V_m^\dag\> \geq
m([qq\la q] - [qq])$.  Next, we will present an 
approximate simulation for $V_m^\dag$ that uses a nearly optimal
amount of communication; i.e. barely more than
$m([qq\la q] - [qq])$.  Thus, by analogy with the nearly optimal
simulation of $V_m$ in \eq{desired-Vm-bounds},
we will have 
\bml
m[qq\la q] + O(\log m/\eps)([q\ra q]+[q\la q]) \\
\gtrsim \<V_m^\dag\> + m[qq] \geq m [qq\la q].
\label{eq:desired-Vmdag}\eml
Again, the first inequality is not really a resource inequality, since
it doesn't give us a way of having the overall error vanish when
simulating $n$ copies of $V_m^\dag$ using $(1+o(1))n$ times the
resource cost.  However, it will still be enough for us to give us
nearly tight bounds on the capacity region of $V_m^\dag$ for $m$ large.

We begin by discussing this capacity region.  Choosing $\eps=1/m$ and
again using the Continuity Lemma, we can show $C_\ra^E(V_m^\dag) \leq
O(\log m) \ll m \leq C_\la^E(V_m^\dag)$, a similar sort of capacity
separation between forward and backwards communication.  However,
$V_m^\dag$ also exhibits a dramatic gap between entanglement-assisted
and -unassisted  capacity.  Since $m[qq\la q] + O(\log m)([q\ra q] +
[q\la q])$ can create no more than $m+O(\log m)$ ebits, the
Continuity Lemma implies that $\<V_m^\dag\> + m[qq]$ also must have
entanglement capacity $\leq m+O(\log m)$.  Finally we use the fact
that the entanglement
capacity of isometries is additive (from \cites{BHLS02,Leifer03} as
well as \thm{C-E-toff}) to establish that
$E(V_m^\dag)\leq O(\log m)$.

This implies that {\em all} unassisted capacities are small; for
example $C_+(V_m^\dag) \leq E(V_m^\dag) \leq O(\log m)$.  Thus
$V_m^\dag$ is almost useless without entanglement; none of its
unassisted capacities are greater than $O(\log  m)$.  On the other
hand, its capacity (from Bob to Alice) rises when entanglement is
supplied, at a rate of nearly one cbit per ebit.  No such behavior is
known for noisy quantum channels, although there are qudit channels
with $O(d)$ multiplicative separations between entanglement-assisted
and unassisted capacities\cites{BSST99,BSST01,Holevo01a}\footnote{Every
example of such a channel has capacity much smaller than $\log d$;
e.g. the channel that maps $\rho$ to $\cN(\rho)=\eps\rho +
(1-\eps)I/d$.  For concreteness, follow \cite{Holevo01a} and choose
$\eps=-1/(d^2-1)$, so $\cN(\rho) = (dI-\rho)/(d^2-1)$.  Then the
single-shot HSW capacity $C^{(1)}(\cN)$ is $\Theta(d^{-3})$ and the
entanglement-assisted capacity $C_E(\cN)$ is $\Theta(d^{-2})$.  Achieving
this entanglement-assisted rate using the protocols of
\cites{BSST01,Holevo01a} requires $\log d$ ebits per use of $\cN$,
meaning that we consume many ebits to get a small enhancement in
classical capacity.  Communication protocols which used less
entanglement were given in \cites{Shor04,DHW05}, but here too we
conjecture that $\Omega(\log d)$ ebits are necessary to raise the
capacity of $\cN$ from $O(d^{-3})$ to $\Omega(d^{-2})$, or even to
$\Omega(d^{-3+\delta})$ for any $\delta>0$.  
}.  We also obtain a separation between
entanglement-creating and -destroying capabilities: $E(V_m^\dag)\leq
O(\log  m)\ll m = E(V_m)$.  (An independently derived, and completely
different, example of a gate with $E(U)\neq E(U^\dag)$ is in
\cite{LSW05}.) 

There are two ways we can derive the simulation of $V_m^\dag$
posited in \eq{desired-Vmdag}.
The simplest is to time-reverse our simulation of $V_m$.  First we
will need to replace the $m[q\ra qq]$ with $\frac{m}{2}([q\ra q] +
[qq])$.  Reversing this will replace $[q\ra q]$ with $[q\la q]$ and
$[qq]$ with $-[qq]$.  Together this is $\frac{m}{2}([q\la q]-[qq]) =
m([qq\la q] - [qq])$, plus of course the $O(\log m/\eps)$ terms.

It may be instructive to also consider a more explicit construction of
the $V_m^\dag$ simulation.  Note that $V_m^\dag$ acts on basis states
$\ket{x}^A\ket{y}^B$ 
as follows:
\ben
V_m^\dag \ket{x}^A\ket{x}^B &=& \ket{x}^A\ket{0}^B
 \qquad\qquad\forall x\\
V_m^\dag \ket{x}^A\ket{y}^B &=& \ket{x}^A\ket{y+1}^B 
\qquad \mbox{for } 0\leq y< x\\
V_m^\dag \ket{x}^A\ket{y}^B &=& \ket{x}^A\ket{y}^B 
\qquad\qquad \mbox{for } y>x
\een
Again Alice and Bob can determine whether $y<x$, $y=x$ or $y>x$ to
accuracy $\eps$ by exchanging $O(\log m/\eps)$ qubits,
and by performing these calculations coherently, can uncompute this
information at the end of the protocol.  As with the $V_m$ simulation,
the three cases at the end of the protocol are different ($0<y\leq x$,
$y=0$ and $y>x$), so it is important that they store no more information
than which case holds.

The interesting case is when $x=y$ and Alice and Bob would like to map
$\ket{x}^A\ket{x}^B \ra \ket{x}^A\ket{0}^B$ for arbitrary values of
$x$.  As we have argued above, this can be simulated by reversing
$m([q\ra qq])=\frac{m}{2}([q\ra q]+[qq])$, which requires a resource
cost of $m([qq\la q]-[qq]) = \frac{m}{2}([q\la q] -[qq])$.  Let us now
examine this reverse procedure in more detail.  For simplicity,
suppose $m=2$.  The procedure we would like to reverse is coherent
super-dense coding ($[q\ra q]+[qq]\geq 2[q\ra qq]$), which maps
$\ket{x_1}^{A_1}\ket{x_2}^{A_2}$ to
$\ket{x_1}^{A_1}\ket{x_2}^{A_2}\ket{x_1}^{B_1}\ket{x_2}^{B_2}$ as
follows: First Alice and Bob add a maximally entangled state
$\ket{\Phi}^{A_3B_1}$.  Then Alice applies the Pauli operator
$X^{x_1}Z^{x_2}$ to the $A_3$ system, leaving the state
\bml\ket{x_1}^{A_1}\ket{x_2}^{A_2}
(X^{x_1}Z^{x_2} \ot I)\ket{\Phi}^{A_3B_1}
=:\\
\ket{x_1,x_2}^{A_1A_2}
\ket{\Phi_{x_1,x_2}}^{A_3B_1},
\eml
where $\ket{\Phi_{x_1,x_2}}:=(X^{x_1}Z^{x_2} \ot I)\ket{\Phi}$.  Note that the
four $\ket{\Phi_x}$ form an orthonormal basis, and thus we can define
the unitary map $U_{\text{SD}} := \sum_{x_1=0}^1\sum_{x_2=0}^1
\ket{x_1,x_2}\bra{\Phi_{x_1,x_2}}$.
Super-dense coding proceeds by Alice sending her half of
$\ket{\Phi_{x_1,x_2}}$ to 
Bob, who applies $U_{\text{SD}}$ to yield the state
$\ket{x_1,x_2}^A\ket{x_1,x_2}^B$. 

Now we explain how this protocol can be reversed to map
$\ket{x_1,x_2}^A\ket{x_1,x_2}^B$ to $\ket{x_1,x_2}^A$.  Bob first
applies $U_{\text{SD}}^\dag$ to $\ket{x_1,x_2}^B$ and obtains
$\ket{\Phi_x}^{B_1B_2}$.  Using $[q\la q]$, Bob sends half of
$\ket{\Phi_x}$ to Alice, so the joint state becomes
$\ket{x_1,x_2}^A\ket{\Phi_{x_1,x_2}}^{AB}$.  Since Alice has a copy of
$x$, she can apply
$(X^{x_1}Z^{x_2})^{\dag}$ to her half of $\ket{\Phi_{x_1,x_2}}^{AB}$ and
transform it to $\ket{\Phi_{0,0}} = \ket{\Phi}$.
Thus, not only has Bob's copy of $\ket{x_1,x_2}$
been erased, but Alice and Bob are left sharing the state
$\ket{\Phi}$.  This means that two bits on Bob's side can be coherently erased
by using the resource $[q\la q]-[qq] = 2([qq\la q] - [qq])$.

\section{Coherent erasure and time-reversal}\label{sec:erasure}
By now we have seen many examples of how unitary communication
protocols can be reversed to give new protocols.  In this section we
summarize these reversal rules and, inspired by our observations about
$V_m^\dag$ in the last section, introduce the new communication
resource of ``coherent erasure,'' which is the time-reversal of
coherent classical communication.  Time-reversal symmetry was also discussed in
\cite{Devetak06}, which explained how the quantum reverse Shannon
theorem\cite{BDHSW08} is the time-reversal of the quantum Slepian-Wolf
theorem (a.k.a. state merging\cite{HOW05} or the
mother\cites{DHW03,DHW05}), once all the protocols are made fully coherent.

For the standard resources of $[q\ra q]$ and $[qq]$, time-reversal is
quite simple.  If we let $^\dag$ denote the time-reverse of a
resource, then $[q\ra q]^\dag = [q\la q]$ and $[qq]^\dag = -[qq]$.
This means that sending qubits in one direction is reversed by sending
qubits in the other direction, and that creating entanglement is
reversed by destroying entanglement.  As we have argued above, $[q\ra
qq] = ([q\ra q] + [qq])/2$ so the time-reversal of $[q\ra qq]$ is
$[q\ra qq]^\dag = ([q\la q] - [qq])/2$.  This resource corresponds to
the map $\ket{x}^A\ket{x}^B\ra \ket{x}^A$, so we call it ``coherent
erasure'' and label it $[q\la qq]$. Since one bit of coherent
classical communication is a cobit, we (following a suggestion of Charlie Bennett's) call $[q\la qq]$ a co-cobit from Bob to Alice, where the
first ``co'' stands for ``complementary'' and the second ``co'' stands
for ``coherent.''  The co-cobit from Alice to Bob is denoted $[qq\ra
q]$ and corresponds to the map $\ket{x}^A\ket{x}^B\ra \ket{x}^B$.  

Of course the map
$\ket{x}^A\ket{x}^B\ra \ket{x}^A$ is not defined for all inputs (what
if Alice and Bob don't input the same state?), but in this section we
will explain how coherent erasure nevertheless makes sense as a
communication resource.  First in \sect{get-coherase} 
we will explain how coherent erasure can be derived from other
resources and then  
in \sect{use-coherase}
we will describe some uses of coherent erasure.

We stress at the outset that coherent erasure is
equivalent to standard resources, and there's no need to introduce new concepts
such as ``erasure capacity'' and the like.  However, it may prove a
useful metaphor in analyzing other communication protocols.

\subsection{ Producing coherent erasure}\label{sec:get-coherase}
We have already seen three ways of producing coherent erasure, which
we briefly review here.
\subsubsection{Reversing clean protocols with $[q\ra qq]$}
If a clean resource inequality involves $[q\ra qq]$, then in the
time-reversed version of the resource inequality $[q\ra qq]$ is
replaced with $[q\ra qq]^\dag = [q\la qq]$.  This was an implicit part
of the argument of \thm{reversal}, which was based on reversing
$\<U\>\geqclean C[q\ra qq]$ to obtain $\<U^\dag\> \geqclean C[q\la
qq]$.
\subsubsection{Reversing super-dense coding} 
This was explained in \sect{Vm-dag}, and is basically a special case
of the last point: time-reversing $[q\ra q] + [qq]\geq 2[q\ra qq]$
yields \be [q\la q] - [qq]\geq 2[q\la qq].
\label{eq:get-coherase}\ee 
\subsubsection{The gate $V_m^\dag$}
Just as $V_m$ is equivalent to $m$ cobits up to small errors and
inefficiencies, $V_m^\dag$ is roughly equivalent to $m$ co-cobits.  Of course $(V_m^\dag)_{m=1}^\infty$ is not a proper
asymptotic resource, but it is still useful as a concrete way to
imagine implementing coherent erasure.

\subsection{Using coherent erasure}\label{sec:use-coherase}
\subsubsection{Entanglement-assisted communication} \thm{reversal} (or
more precisely, the protocol sketched in \sect{reversal} for
entanglement-assisted back
communication using $\Uxoxo$)
explained how
\be [q\la qq] + [qq] \geq [qq\la q] 
\label{eq:use-coherase}\ee
 by giving an explicit protocol.  Another way to derive this resource
inequality is reversing coherent teleportation 
$$2[q\ra qq] \geq [q\ra q]+[qq]$$
to obtain 
$$2[q\la qq] \geq [q\la q] - [qq].$$
Substituting $[q\la q] = 2[qq\la q] - [qq]$ then gives the desired
result.

Interestingly, coherent erasure is capable of no communication on its
own, but can convert one ebit into one cobit.  This is a sharper
version of the separation we observed between the
entanglement-assisted and -unassisted capacities of $V_m^\dag$.

\subsubsection{State merging and partial quantum communication}
First note that \eqs{get-coherase}{use-coherase} can be combined to
obtain the equality
\be\begin{split} [q\la qq] &= [qq\la q] - [qq] 
\\&= \frac{[q\la q] - [qq]}{2}
\\& = [q\la q] - [qq\la q].
\label{eq:coherase-eq}\end{split}\ee
This last point means that $[q\ra q] = [q\ra qq] + [qq\ra q]$, so the
task of sending a qubit can be split into the tasks of sending one
cobit and coherently erasing one bit.  There is a direct protocol
which performs this.  Suppose Alice would like to send the state $\sum_x
\alpha_x \ket{x}^A$ to Bob.  If she applies $[q\ra qq]$ then they will
obtain the entangled state $\sum_x \alpha_x \ket{x}^A\ket{x}^B$.  Finally,
applying $[qq\ra q]$ will erase Alice's state and leave Bob with
$\sum_x \alpha_x \ket{x}^B$.

Of course this also works for coherent superpositions of messages.  If
 Alice would like to send her half of $\sum_x\alpha_x
 \ket{x}^R\ket{x}^A$ to Bob then she can first apply $[q\ra qq]$ to
 obtain 
$$\sum_x \alpha_x \ket{x}^R\ket{x}^A\ket{x}^B,$$ 
and then $[qq\ra q]$ will again erase Alice's state to leave
$\sum_x \alpha_x \ket{x}^R\ket{x}^B$.

The general problem here is {\em state merging}\cite{HOW05}, in which
Alice gives Bob her piece of a tripartite state  $\ket{\psi}^{ABR}$,
perhaps generating or consuming entanglement in the process.  
\mscite{Devetak06} argued that cobits are the canonical example of
feedback channels, which are isometries that map from $A$ to $AB$.
Likewise, we claim that coherent erasure is the canonical example of
state merging.

To justify this interpretation, we will now show how to generalize the
decomposition $[q\ra q] = [q\ra qq] + [qq\ra q]$ to a decomposition of
perfect quantum communication from $A\ra B$ into an
isometry from $A\ra AB$ followed by  merging $AB$ into $B$.   The
isometry from $A\ra AB$ can be simulated by $I(R;B)[q\ra qq] +
I(B\> A)[qq]$ using the quantum reverse Shannon
theorem\cites{BDHSW08,Devetak06}, where the coherent information $I(B\>
A)$ is defined as  $I(B\> A):= -H(B|A) =
H(A)-H(AB)=H(A)-H(R)$.  The $I(R;B)[q\ra qq]$ cost represents the
difficulty of creating the desired correlations between Bob and the
reference system, while the $I(B\> A)[qq]$ cost is necessary because
back communication would allow Bob to distill that much entanglement
with Alice while preserving his correlations with $R$.
Then the tripartite state can be mapped (using state merging) to one
where Bob holds Alice's 
part using $I(R;A)[qq\ra q] - I(A\> B)[qq]$.  Here the erasure cost
measures the amount of correlation with the reference system that
Alice has and needs to give up, while $I(A\>B)$ is the amount of
entanglement that is recovered by the procedure once Bob has the
entire purification of Alice's state.  Indeed, this version
of state merging amounts to a coherent version of entanglement
distillation ($I(A;E)[c\ra c] + \<\psi^{ABE}\> \geq I(A\>B)[qq]$) in
which $[c\ra c]$ is replaced with $[qq\ra q]$ and as a result Bob
holds the purification of the environment at the end of the protocol.
Finally, the total resource cost 
$I(R;B)[q\ra qq] + I(B\> A)[qq] + I(R;A)[qq\ra q] - I(A\> B)[qq]$ is
simply equal to $H(R)[q\ra q]$, the cost of sending the reference
system directly to Bob.

\subsubsection{Rule I: Coherently decoupled input cbits}
Suppose $\alpha + C [c \ra c] \geq \beta$ is a resource inequality in
which the classical message sent is nearly
independent of all residual quantum systems, including the
environment.  In this case we say that the input cbits are {\em coherent
decoupled} and ``Rule I'' of \cites{DHW03, Har05,DHW05} proves that
they can be replaced by 
$C([q\ra qq] - [qq])$.  Equivalently, we can replace them by $C[qq\ra
q]$.  Thus, coherent erasure can be used whenever we need to send a
classical message whose contents can be guaranteed to be almost
completely independent of the remaining quantum systems.

The simplest example of a coherently decoupled input is of course
teleportation, which quickly leads us to the familiar resource
inequality $2[qq\ra q] + [qq] \geq [q\ra q]$.  A slightly more
nontrivial example is remote state preparation\cite{BHLSW03}, in which
$\log d$ ebits and $\log n:=\log d + \log \log d + 2\log 1/\eps + O(1)$
cbits are used for Alice to prepare an arbitrary $d$-dimensional state
in Bob's lab (i.e. $\log d$ ``remote qubits'').  Since the input cbits
are coherently decoupled, they can be replaced by $(\log n)([q\ra
qq]-[qq])$.  Asymptotically this means that one coherent bit is at
least as strong as one remote qubit, which is an interesting statement
because cobits and remote qubits both lie somewhere in between cbits
and qubits, and there does not appear to be any trivial protocol
relating the two.  Another way to interpret remote state preparation
is that $[qq\ra q] + [qq]$ yield one remote qubit; this, by contrast,
can be implemented by a relatively straightforward single-shot
protocol.  Suppose the state that Alice wishes to prepare for Bob is
$\ket{\alpha}=\sum_{x=1}^d \alpha_x \ket{x}$.  A particularly easy
case is when 
$|\alpha_x|^2 = 1/d$ 
for all $x$.  Then Alice could locally map $\ket{\Phi_d}:= 
\frac{1}{\sqrt{d}}\sum_{x=1}^d  \ket{x}^A\ket{x}^B$ to $\sum_{x=1}^d
\alpha_x \ket{x}^A\ket{x}^B$ and then use $\log d[qq\ra q]$ to leave
Bob with the state $\sum_{x=1}^d\alpha_x \ket{x}^B$.

In general, $|\alpha_x|$ will not always be equal to $1/\sqrt{d}$, and
blithely applying the above method for a general state will only
achieve a fidelity of $F(\alpha):=\sum_x |\alpha_x|/\sqrt{d}$.  Moreover,
even a randomly chosen state will usually have $F$ close to
$\sqrt{\pi}/2$ when $d$ is large (as can be seen using
$\bbE|\alpha_x|=\sqrt{\pi/4d}$ and the central limit theorem).  This
means that multiplying $\ket{\alpha}$ by a fixed random unitary will
not be sufficient to obtain high fidelity.  Instead, we will use a
small sequence of unitaries $\{U_1,\ldots,U_\kappa\}$, along with a
$\kappa$-dimensional ancilla register that 
controls which unitary is applied.  To decode, Bob will need this
ancilla register, which we will require $\log\kappa$ extra qubits of
communication from Alice to Bob.  Fortunately, we will see that the
error shrinks rapidly with
$\kappa$ for a variety of choices of $\{U_1,\ldots,U_\kappa\}$.

The procedure for Alice to remotely prepare $\ket{\alpha}$ in Bob's
lab is as follows.  Let
$$\ket{\beta}:=\frac{1}{\kappa}\sum_{k=1}^{\kappa} \ket{k}
U_k\ket{\alpha} = \sum_{x=1}^d \beta_x \ket{b_x}\ket{x},$$
where in the last expression we have introduced states $\ket{b_x}$ for
each $x$.  Assume for now that $F(\beta)=\sum_x |\beta_x|/\sqrt{d}$ is
close to one.  Then, starting with the shared state $\ket{\Phi_d}$,
Alice can create 
$$\frac{1}{\sqrt{d}}\sum_x
\ket{x}^A\ket{x}^B\ket{b_x}^{A'}.$$
If she then applies $\log d$ co-cobits to $AB$ and sends $A'$ to Bob
using $\log \kappa$ qubits then Bob will have a fidelity-$F(\beta)$
approximation to $\ket{\beta}$, from which he can obtain
$\ket{\alpha}$.

We can summarize the correctness of the protocol as follows:
\begin{theorem}
There exists a subspace $V\subset \bbC^d$ with $d':=\dim V =
d\eps^2/32\log(5/\eps)$ such that the remote state preparation
protocol above can prepare any state $\ket{\alpha}\in V$ in Bob's lab
using $\log d [qq\ra q]+\log d [qq] +\log \kappa[q\ra q]$ and with
fidelity $1-\frac{1}{2\kappa}-2\eps$.
\end{theorem}

This translates into remotely preparing a state of $\log d'= \log d -
2\log 1/\eps - O(\log\log 1/\eps)$ qubits, which is similar to the
performance of \cite{BHLSW03}.

\begin{proof}
To analyze the protocol, we first seek to lower bound $\bbE_\alpha
F(\beta)$, where $\alpha$ is uniformly randomly chosen.  We will
choose $U_k$ to be $\sum_x \ket{x+k}\bra{x}$ for
$k\in\{1,\ldots,\kappa\}$ and where addition is mod $d$.  We could
also choose the $U_k$ to be (approximately) mutually unbiased bases
(meaning that $|\bra{x}U_k^\dag U_{k'}\ket{x}|^2$ is (approximately)
equal to $1/d$), but we omit the analysis here.  To evaluate the
expected fidelity, 
we use linearity of expectation:
\be\begin{split}\bbE_\alpha F(\beta) &=
\frac{1}{\sqrt{d}} \sum_{x=1}^d \bbE_\alpha |\beta_x| 
\\&=
\frac{1}{\sqrt{d\kappa}} \sum_{x=1}^d
\bbE_\alpha \sqrt{\sum_{k=1}^\kappa |\braket{x+k}{\alpha}|^2}.
\end{split}\ee
Using the fact that the distribution of $\ket{\alpha}$ is rotationally
invariant, we find that
$$\bbE_\alpha F(\beta) = \sqrt{\frac{d}{\kappa}} \bbE_\alpha 
\sqrt{\tr P\alpha},$$
where $P$ is a rank-$\kappa$ projector (assuming that $\kappa\leq
d$). To estimate this last term, we 
 will use
the inequality $\bbE |X| \geq
(\bbE X^2)^{\frac{3}{2}} / (\bbE X^4)^{\frac{1}{2}}$ which holds for any
random variable~\cite{Berger91}.  Next, we calculate $\bbE \tr P\alpha =
\kappa/d$ and $\bbE (\tr P\alpha)^2 = 
\tr (P\otimes P)\cdot \bbE (\alpha\otimes \alpha) = \tr (P\otimes
P)\cdot(I+\textsc{SWAP})/d(d+1)=\kappa(\kappa+1)/d(d+1)$.  Putting
this together we find that 
$$\bbE_\alpha F(\beta) \geq
\sqrt{\frac{1+\frac{1}{d}}{1+\frac{1}{\kappa}}}
\geq 1- \frac{1}{2\kappa}.$$

The remaining steps are quite similar to the arguments in
\cites{AHSW04, HLW06}.  We will argue that not only is $F$ close to its
expectation for most values of $\alpha$, but in fact if we choose a
random subspace $V$ of dimension $d' = O(d\eps^2 
/32\log(5/\eps))$ then with nonzero probability every vector
$\ket{\alpha}\in V$ will have $F(\beta)\geq
1-\frac{1}{2\kappa}-\eps$.   Observe that the Lipschitz constant of
$F(\beta)$ (defined to be $\max_{\ket{\alpha}} \sum_x
(\partial F/\partial_{\alpha_x})^2 $) is constant.  In fact, it is 1.
Then Levy's Lemma\cite{Ledoux01} states that
\be \Pr_\alpha\l[F(\alpha)\leq 1-\frac{1}{2\kappa}-\eps\r] \leq
2 \exp\l(-\frac{d\eps^2}{31}\r)
\label{eq:one-pt-good},\ee
for any $\eps>0$.
Now consider a random $d'$-dimensional subspace $V$.  According to
\cite{HLW06}*{Lemma III.6}, we can choose a
mesh of $(5/\eps)^{d'}=\exp(d\eps^2/32)$ points such that 
any point in $V$ is within $\eps$ (in trace distance) of some point in
the mesh.  Applying 
the union bound and the triangle inequality to \eq{one-pt-good}, we
find that there exists a subspace $V$ such that $F(\alpha)\geq
1-\frac{1}{2\kappa}-2\eps$ for all $\ket{\alpha}\in V$.
\end{proof}

Thus coherent erasure is an alternate, and arguably more direct, way
to think about remote state preparation.  It is perhaps interesting
that our protocol is nontrivially different from the comparably
efficient protocol that could be obtained from applying
\cites{DHW03,Har05,DHW05}'s ``Rule I'' to \cite{BHLSW03}.  While Rule
I guarantees that in general co-cobits can be used in place of
coherently decoupled input bits, the proof is indirect and involves
catalysis.  It would be interesting to know whether there is a more
direct and natural use of coherent erasure, such as the one we showed for
remote state preparation, in any protocol with coherently decoupled
input cbits.

\section{Conclusions}\label{sec:conclude}
The results of this paper help resolve many questions surrounding
unitary gate capacities.  We now understand that, while all the
capacities of any nonlocal
gate are nonzero, there can be asymptotically large
separations between these capacities.  The main separation left unproven by this paper is finding a gate with $E(U) \gg C_+(U)$.  An early version of this paper proposed the gate on $\bbC^{d}\ot\bbC^{d}$
which exchanges $\ket{01}$ and $\ket{\Phi_d}$, while leaving the other
states unchanged: $U = I - |01\>\<01| - |\Phi_d\>\<\Phi_d| +
|01\>\<\Phi_d| + |\Phi_d\>\<01|$.  Since $U$ requires $\log d$ cbits to simulate, even using unlimited EPR pairs (as we will prove later in this section), the simulation techniques in this paper will need to be modified.  Since the first version of this paper appeared, \cite{HL07} established the conjectured separation (showing that $C_+^E(U)\leq O(\log\log d) \ll \log d  \leq E(U)$) by using non-maximally entangled states for the simulation.

Another limitation of our work is that the separations we have
found are 
between $o(\log d)$ and $\approx \log d$; on the other
hand, \cite{LSW05} has proven that if $E(U)=2\log d$ then
$E(U^\dag)=2\log d$ as 
well.  It would be interesting to see which capacity separations are
possible for gates with $E(U)$ between $\log d$ and $2\log d$.

Of course, separating communication capacities is mostly intended as a
step towards better understanding quantum communication using unitary
gates as well as other resources.  For example, it has led us to the
resource of coherent erasure, which hopefully will turn out to be a 
useful concept the way coherent classical communication has.

Entanglement destruction is another resource in quantum information
theory that seems to be worth exploring.  Once we restrict protocols
to be clean, destroying entanglement is a nonlocal task.
Equivalently, creating coherent superpositions of states with varying
amount of entanglement requires communication, even if the two
parties are allowed unlimited numbers of maximally entangled
states\cite{HW02}.  This task comes up in entanglement
dilution\cites{HW02,HL02}, in its generalization, remote
preparation of known entangled states (though this is not
explicitly acknowledged in \cite{BHLSW03}), 
and in the quantum reverse Shannon theorem\cite{BDHSW08}, which may be
thought of as a further generalization of remote state preparation.
In each case, the resource of ``entanglement spread''---meaning the
ability to generate superpositions of states with varying amounts of
entanglement---appears to be necessary.  Entanglement spread can be
generated (using e.g. remote state preparation) by sending cbits in
either direction, or even without any communication at all, if Alice
and Bob can make catalytic use of an embezzling state\cite{HV03}.
It appears that a unitary gate $U$ has ``spread capacity'' equal to
$E(U)+E(U^\dag)$; for example, a simple application of \cite{HW02} can
prove that $C_1[c\ra c]+C_2[c\la c]+\infty[qq]\leq \<U\>$ implies that
$C_1 + C_2 \geq E(U) + E(U^\dag)$, a result previously proved only for
two-qubit gates, using very different arguments~\cite{BS03b}.  Of
course, without a precise 
definition of 
entanglement spread it will be difficult to formalize these arguments.
\mscite{HW02} is a promising first step towards
defining spread as a resource, though the unusual scaling of
embezzling states and of the cost of entanglement dilution suggest
that the i.i.d. resource model of \cites{DHW05,Har05} might not fit
well. 

\section{Notation}\label{sec:notation}
In this section, we collect some of the notation used in the rest of the
paper.
\begin{table}[!h]
\renewcommand{\arraystretch}{1.6}
\caption{Definitions of communication resources}
\label{tab:cq-def}
\centering
\begin{tabular}{|c|c|c|}\hline
 abbr. & name & formula \\ \hline 
$[qq]$ & ebit & $\frac{1}{\sqrt 2}(\ket{00}^{AB}+\ket{11}^{AB})$
\\ \hline
$[c\ra c]$ & cbit & 
$\sum_{x=0}^1 \ket{x}^B\ket{x}^E \bra{x}^A$  \\ \hline
$[q\ra q]$ & qubit & 
$\sum_{x=0}^1 \ket{x}^B \bra{x}^A$  \\ \hline
$[q\ra qq]$ & cobit & 
$\sum_{x=0}^1 \ket{x}^A\ket{x}^B \bra{x}^A$ \\ \hline
$[q\la qq]$ & co-cobit & 
$\sum_{x=0}^1 \ket{x}^A\bra{x}^A \bra{x}^B$ \\ \hline
$\langle U \rangle$ & unitary & $U$ \\ \hline
\end{tabular}
\end{table}

Cobits were introduced in \cite{Har03} and co-cbits were introduced in
\sect{erasure}.  Note that co-cobits are only defined when Alice and
Bob's joint state is constrained to lie in the subspace spanned by
$\ket{00}$ and $\ket{11}$.

To understand the relations between the resources in \tabref{cq-def},
we describe the 
effects of exchanging Alice and Bob and of running a protocol
backwards.  These are listed in the ``exchange'' and ``reverse''
columns of the table below.  For example, $[q\ra q]$ means sending a
qubit from Alice to Bob, so either exchanging Alice and Bob
or reversing time transforms $[q\ra q]$ to $[q\la q]$.  However,
given a cobit from Alice to Bob, exchange and time-reversal do not act
the same way: exchanging Alice and Bob yields a cobit from Bob to
Alice while time-reversal yields a co-cobit from Bob to Alice.

\begin{table}[!h]
\renewcommand{\arraystretch}{1.3}
\caption{Exchange and time-reversal symmetries}
\label{tab:cq-sym}
\centering
\begin{tabular}{|c|c|c|}\hline
resource & exchange & reverse \\ \hline 
$[qq]$ & $[qq]$ & $-[qq]$ \\ \hline
$[c\ra c]$ & $[c\la c]$ & undefined \\ \hline
$[q\ra q]$ & $[q\la q]$ & $[q\la q]$ \\ \hline
$[q\ra qq]$ & $[qq\la q]$ & $[q\la qq]$ \\ \hline
$[q\la qq]$ & $[qq \ra q]$ & $[q\ra qq]$ \\ \hline
$\<U\> $ & $\< \cF U \cF\>$ & $\<U^\dag \>$ \\ \hline
\end{tabular}
\end{table}

In the first line of \tabref{cq-sym}, we can consider $[qq]$ to be the
action of creating an ebit. Thus, the time-reversal of $[qq]$ corresponds to
destroying an ebit, which, if done coherently, is a non-trivial
resource.  In the last line, $\cF$ denotes the unitary operator that
exchanges Alice and Bob's systems.  The time-reversal relations for
cobits and co-cobits are explained in \sect{erasure}.

Next, we summarize some of the basic transformations of the resources
in \tabref{cq-def} that are possible.  We have indicated where the
protocols can be done cleanly by using $\geqclean$ or $\eqclean$
instead of $\geq$ or $=$.

\begin{table}[!h]
\caption{Transformations between standard resources}
\label{tab:cq-proto}
\vspace{-8mm}
\begin{align*}
2[c\ra c] + [qq] & \geqclean [q\ra q] & \text{teleportation} \\
[q\ra q] + [qq] & \geqclean 2[c\ra c] & \text{super-dense coding}\\
[q\ra q] + [qq] & \eqclean 2[q\ra qq] & \text{from \cite{Har03}}
\\
[q\ra q] - [qq] & \eqclean 2[q\la qq] & \text{from \sect{erasure}}
\\
[q\ra qq] + [q\la qq] & \eqclean [q\ra q] &\text{from last two lines} 
\end{align*}
\end{table}

Now define $U$ to be a bipartite unitary gate.  We have defined
various capacity regions in \sect{intro}, which are summarized in
\tabref{cap-regions} below.
\begin{table}[!h]
\caption{Capacities and capacity regions of unitary gates}
\label{tab:cap-regions}
\begin{align*}
\CCE(U) & = 
\{(C_1,C_2,E) : U \geq C_1 [c\!\ra\! c] + C_2 [c\!\la\! c] + \! E[qq] \}\\
\clCCE(U) & = 
\{(C_1,C_2,E) : U \!\!\geqclean\!\! 
C_1 [c\!\ra\! c] + C_2 [c\!\la\! c] +\! E[qq] \}\\
\CE(U) & = \{(C,E) : U \geq C [c\ra c] +  E[qq] \}\\
\clCE(U) & = \{(C,E) : U \geqclean C [c\ra c] + E[qq] \}\\
C_\ra^E(U) & = \max \{ C : U + \infty [qq] \geq C [c\ra c]\} \\
C_\ra(U) & = \max \{ C : U \geq C [c\ra c]\} \\
C_\la^E(U) & = \max \{ C : U + \infty [qq] \geq C [c\la c]\} \\
C_\la(U) & = \max \{ C : U \geq C [c\la c]\} \\
C_+(U) & = \max \{C_1 + C_2 : U \geq C_1 [c\ra c] + C_2 [c\la c]\} \\ 
C_+^E(U) & = \max \{C_1\!+\! C_2 : U \!+\! \infty[qq]
\geq C_1 [c\!\ra\! c] + C_2 [c\!\la\! c]\} \\ 
E(U) & = \max\{E : U \geq E [qq] \} 
\end{align*}
\end{table}

Finally, we will explain how the results of the paper relate to the terms in  \tabref{cap-regions}.  First, some of the theorems relate as follows:
\bitem \item \thm{std-form} shows that the capacity regions $\CCE$ and $\clCCE$ are nearly equivalent, other than the fact that entanglement can be thrown away in non-clean protocols.  As a corollary, \thm{std-form} also relates $\CE$ to $\clCE$ the same way.
\item \thm{C-E-toff} and \lem{continuity} concern the region $\clCE(U)$, giving a single-letter formula for it, and proving its continuity (in terms of $U$), respectively.
\item \thm{reversal} shows that $\clCCE(U^\dag)$ can be completely determined from $\clCCE(U)$ (and vice-versa, of course).
\eitem
Finally, \tabref{separations} summarizes the separations in capacities proved in \sect{Vm} and \sect{Vm-dag}. 

\begin{table}[!h]
\renewcommand{\arraystretch}{1.5}
\caption{Exchange and time-reversal symmetries}
\label{tab:separations}
{\centering
\begin{tabular}{|c|c|} \hline
f. vs. b. comm. & 
$C_\la^E(V_m) \ll C_\ra(V_m)$ \\ \hline
ent. create/destroy & 
$E(V_m^\dag) \ll E(V_m)$  \\ \hline
ent.-assted f.~vs.~b.~comm. &
$C_\ra^E(V_m^\dag) \ll C_\la^E(V_m^\dag)$ \\ \hline
comm. w/ or w/o ent.-asst. & 
$C_+(V_m^\dag) \ll C_+^E(V_m)$ \\ \hline
\end{tabular}

\medskip Here f.~means forward, b.~means backwards, comm.~means communication (which can be equivalently taken to be cbits, cobits or qubits), ent.~means entanglement, and asst/assted means assistance/assisted. }

\end{table}

\section*{Acknowledgments}
 We would like to thank Andreas Winter for
allowing us to include \thm{winter-concentrate}, Harry Buhrman for
teling us about \cite{Nisan93} and Noah Linden for
telling us about \cite{LSW05}.  AWH would also like to thank Charlie
Bennett, Debbie Leung and John Smolin for suggesting the problem of
asymmetric unitary gate capacities and for many interesting
discussions on the subject.

\appendix
\section{New proofs of Theorems~\ref{thm:std-form} and
\ref{thm:C-E-toff}}\label{app:reprove}
In this appendix we sketch alternate proofs for Theorems
\ref{thm:std-form} and \ref{thm:C-E-toff}.  Both proofs we give are slightly
simpler than previous versions and have slightly better convergence
properties.  Moreover, by including them, this paper can be more
self-contained, especially given that the previous statements of
\thm{std-form} in \cites{HL04,Har05} were slightly weaker.

\myproof{of \thm{std-form}} We begin by following the approach of
\cite{HL04}, where this was first proved.  Suppose $\<U\> \geq C_1
[c\ra c] + C_2 [c\la c] + E [qq]$ for some $E>0$ (similar arguments
apply for $E\leq 0$).  Then, if Alice and Bob copy their inputs before sending and
refrain from performing their final von Neumann measurements, we
obtain a sequence of protocols
$\cP_n$ such that for all $x\in\{0,1\}^{\Cna}$
and $y\in\{0,1\}^{\Cnb}$ (with $C_j^{(n)} := n(C_j - \delta_n)$ for
$j=1,2$)
\bml
\cP_n \ket{x}^A \ket{y}^B \upto{\epsilon_n}\\
\ket{x,y}^A \ket{x,y}^B (\ket{\Phi}^{AB})^{\ot n(E-\delta_n)}
\ket{\varphi_{x,y}}^{AB}.
\label{eq:std-form3}\eml
The main difference between \eq{std-form3} and our goal
(\eq{std-form2}) is the presence of the ancilla
$\ket{\varphi_{x,y}}^{AB}$ with its arbitrary depends on $x$ and $y$
rather than a string of zeroes that depends only on $n$.  Simply
discarding $\ket{\varphi_{x,y}}^{AB}$ will in general break
superpositions between different values of $x$ and $y$.  Also the
ancilla is not guaranteed to fit in $o(n)$ qubits.

\mscite{HL04} made a series
of modifications in order 
to obtain a clean protocol.  In fact, \cite{HL04} obtained a slightly
weaker result than \eq{std-form2}, in which Alice and Bob are left
with an ancilla $\ket{\varphi_n}^{AB}$ which depends only on $n$ and
(it can be shown) can be stored in $\leq n\delta_n'$ qubits.  We call
protocols of this form ``semi-clean,'' but when working with unitary
gates this implies the protocol can be made clean at an asymptotically
negligible additional cost.  This is due to \cite{Har08a}, which
proved that any nonlocal gate $U$ can exactly generate any other fixed
gate, such as \textsc{swap}, with a constant number of applications
interspersed with local unitaries.  Thus, $O(n\delta'_n)$ applications
of $U$ can exactly map $\ket{\varphi_n'}$ to $\ket{00}^{n\delta_n'}$.

We now review the steps of \cite{HL04} in obtaining a semi-clean
protocol before we describe our alternate approach.  First they used
entanglement as a sort of coherent one-time-pad, so that the ancillas
would become correlated with the one-time-pad and not the message.
Then a classical error-correcting code was applied to reduce the errors, and
finally entanglement concentration\cite{BBPS96} was used on the
error-free blocks to recover the entanglement used for the
one-time-pad.  We use an approach that is only slightly
different: first using block coding to reduce the error to a nearly
exponentially small amount, then using a coherent one-time-pad to
decouple the ancillas and finally recovering entanglement using an
approximate form of entanglement concentration that needs far fewer
states.

We now explain these components in more detail.
First, absorb all of
the output into the ancilla, except for the locally copied inputs, so that
\be
\cP_n \ket{x}^A \ket{y}^B =
\ket{x}^{A_1}\ket{y}^{B_1}
\ket{\varphi_{x,y}}^{A_2A_3B_2B_3},
\label{eq:std-form4}
\ee
with $\tr(\proj{y}^{A_2}\ot \proj{x}^{B_2} \ot
I^{A_3B_3})\varphi_{x,y} \geq 1-\eps_n$ and 
with entanglement $E(\ket{\varphi_{x,y}}) := H(\varphi_{x,y}^{A_2A_3})
\geq n(E-\delta_n)$, for an appropriate 
redefinition of $\delta_n$.
This is clearly an equivalent formulation, but it allows us to speak
more easily about the exact output of the protocol.

Now we will use 
classical bidirectional block-coding\cite{Shannon61} 
to control how quickly $\eps_n$ vanishes as a function of $n$.  
By applying
$\cP_{n_1}$ $n_2$ times and slightly reducing the rate, it is possible
to send
$n_1n_2(C_1-\delta)$ cbits forward and
$n_1n_2(C_2-\delta)$ cbits backwards with $n_1n_2$ uses of
  $U$ and average error $\leq\exp(n_2(1+\alpha\log\eps_n))$ as long as $\delta
\geq \delta_{n_1} + \alpha +
H_2(\alpha)/(n_1(\min(C_1,C_2)-\delta_{n_1}))$.  This can be
simplified by choosing $\alpha = 2/\log(1/\eps_n)$, so the error 
probability is $\leq \exp(-n_2)$ and we still have $\delta\ra 0$.  Finally, for any $0<\gamma<1$,
choose $n_2=n_1^{\frac{1-\gamma}{\gamma}}$, so if $n:=n_1n_2$, then
$n_2 = n^{1-\gamma}$.  Thus, we can without loss of generality assume
that we have a sequence of protocols $(\cP_n)_{n=1}^\infty$ with
inefficiency 
$\delta_n$ and error $\eps_n$ satisfying $\lim_{n\ra
\infty}\delta_n=0$ and $\eps_n \leq
\exp(-n^{1-\gamma})$ (in fact, we could choose $\eps_n\leq\exp(-f(n))$
for any $f(n)=o(n)$ that we like).\footnote{This step closely
resembles the code 
construction in \cite{HL04}, which contains a slightly more detailed
proof.  The basic tools for the proof can also be found in
\cites{CT91,Shannon61}; note that they (esp. \cite{Shannon61}) only
bound the {\em average} decoding error, rather than the maximum
error (similarly in \cite{BS03b}, which introduced the idea of
double-blocking unitary communication protocols).  Average error can easily
be turned into maximum error\cite{DW05}, though in our case
it is not necessary, or rather, our entire protocol can be thought of
as a coherent version of \cite{DW05}.}  Now we see the reason for
using the form of \eq{std-form4}; our block codes still satisfy
\eq{std-form4}, but not necessarily \eq{std-form3}, since there may be
arbitrary errors in the entangled states.

 Next, use a coherent one-time-pad in the same way as
\cite{HL04}.  Alice and Bob start with $\Cna +
\Cnb$ ebits.  Together with their initial messages
$\ket{x}^A\ket{y}^B$, their state is
$$\frac{1}{\sqrt{N}}
\sum_{a\in\cA}
\sum_{b\in\cB}
\ket{a,b,x}^A \ket{a,b,y}^B,
$$
where $\cA:=\{0,1\}^{\Cna}$, $\cB:=\{0,1\}^{\Cnb}$, and
$N := |\cA|\cdot |\cB| = \exp(\Cna + \Cnb)$.  This can be locally mapped
to
$$\frac{1}{\sqrt{N}}
\sum_{a\in\cA}\sum_{b\in\cB}
\ket{a,b,a\oplus x}^A \ket{a,b,b \oplus y}^B.
$$
Relabelling the sum over $a,b$ shows that this is equivalent to
$$\frac{1}{\sqrt{N}}
\sum_{a\in\cA}
\sum_{b\in\cB}
\ket{a\oplus x,b\oplus y,a}^A \ket{a\oplus x,b\oplus y,b}^B.
$$
Now $\cP_n$ is applied to $\ket{a}^A\ket{b}^B$, obtaining
$$\frac{1}{\sqrt{N}}
\sum_{a\in\cA}
\sum_{b\in\cB}
\ket{a\oplus x,b\oplus y,a}^A \ket{a\oplus x,b\oplus y,b}^B
\ket{\varphi_{a,b}}.
$$
Since $\tr(\proj{b}^{A}\ot \proj{a}^{B} \ot
I^{AB})\varphi_{a,b} \geq 1-\eps_n$, we can extract
$\ket{b}^A\ket{a}^B$ from $\ket{\varphi_{a,b}}$ while causing
$O(\sqrt{\eps_n})$ disturbance.  This yields a state within
$O(\sqrt{\eps_n})$ of 
$$\frac{1}{\sqrt{N}}
\sum_{\substack{a\in\cA\\b\in\cB}}
\ket{a\oplus x,b\oplus y,a,b}^A \ket{a\oplus x,b\oplus y,b,a}^B
\ket{\varphi_{a,b}},
$$
which can be locally mapped to
\begin{multline*}\ket{x,y}^A\ket{x,y}^B
\frac{1}{\sqrt{N}}
\sum_{a\in\cA}\sum_{b\in\cB}
\ket{a,b}^A \ket{a,b}^B\ket{\varphi_{a,b}}
\\=: \ket{x,y}^A\ket{x,y}^B \ket{\bar{\varphi}}^{AB}.
\end{multline*}
Thus, we have performed the desired coherent communication (up to
error $O(\sqrt{\eps_n}) = O(\exp(-n^{1-\gamma}/2))$) and converted
$n(C_1 + C_2 - 2\delta_n)$ ebits into $\ket{\bar{\varphi}}^{AB}$,
which is the fixed pure state
$$\frac{1}{\sqrt{N}}
\sum_{a\in\cA}
\sum_{b\in\cB}
\ket{a,b}^A \ket{a,b}^B\ket{\varphi_{a,b}}.$$

The final step is to recover entanglement from
$\ket{\bar{\varphi}}^{AB}$.  Since $E(\varphi_{a,b})\geq
n(E-\delta_n)$, we can bound $E_0 := E(\bar{\varphi})\geq n(C_1 + C_2
+ E - 3\delta_n)$.  Also, $\ket{\bar{\varphi}}^{AB}$ can be obtained
from $O(n)[qq] + O(n)\<U\>$, so it must have Schmidt rank $\leq
\exp(O(n))$.  We would like to repeat the entire
protocol $k$ times in parallel, and then apply entanglement
concentration to $\ket{\bar{\varphi}}^{\ot k}$ in order to recover
standard EPR pairs.  This will inevitably increase our error; we get
up to $k\cdot O(\sqrt{\epsilon_n})$ from repeating the protocol, and
some additional errors from the entanglement concentration.  However,
as long as our final error $\eps'$ is $o(1)$ then we can use the
catalytic entanglement safely; we merely repeat the protocol $\lceil
1/\eps' \rceil$ times for a total error of $\sqrt{\eps'}$ and an
additional fractional inefficiency of $\sqrt{\eps'}$, both of which are
still $o(1)$.

Unfortunately, the original entanglement concentration protocol of
\cite{BBPS96} will not suffice for this purpose (without using a more
elaborate procedure, as in \cite{HL04}).  This is because
\cite{BBPS96} requires 
$k / \log k$ to grow faster than the Schmidt rank of
$\ket{\bar{\varphi}}$, meaning that $k=\exp(\Omega(n))$.  Since
$\eps_n = \exp(-o(n))$, we would be unable to guarantee that $k\eps_n
= o(1)$.

Instead, we will use an approximate version of entanglement
concentration (due to Andreas Winter\cite{Winter05}) which only
requires $k=\poly\log(\sch(\bar{\varphi})) = \poly(n)$ to achieve
vanishing error and inefficiency.  Since $\poly(n)\eps_n=o(1)$, this
will complete our proof.
\endproof

It now remains only to describe Winter's new version of entanglement
concentration\cite{Winter05}.  Since this result may be more broadly
useful, we rename the variables from the proof of \thm{std-form} to
more conventional notation, and state a slightly more general result
than we need above.

\begin{theorem}[due to A. Winter]\label{thm:winter-concentrate}
Let $\ket{\psi_1}^{AB},\ldots,\ket{\psi_n}^{AB}$ be bipartite states
each with Schmidt rank $\leq d$, and with total entanglement $E :=
\sum_{i=1}^n H(\psi_i^A)$.  Then for any
$\delta>0$ such that $n \geq \max(\delta^{-2}3(\log d)^3,
20\delta^{-1}\log n\delta)$, Alice and Bob 
can extract $E-n\delta$ ebits from
$\ket{\psi_1}^{AB}\ot\ldots\ot\ket{\psi_n}^{AB}$ with error
$O(\exp(-(n\delta^2)^{1/3}/2\ln 2))$ using no
communication.   Up to the above error, their residual state has
Schmidt rank $\leq \exp(2n\delta)$.
\end{theorem}
In particular, 
$E(1-o(1))$ ebits can be extracted with error $o(1)$ while creating a
sublinear-size garbage state if
$n=\poly(\log d)$.  This suffices to complete the above proof
of \thm{std-form}.

\myproof{of \thm{winter-concentrate}}
First we describe the protocol, then analyze its correctness.
Alice and Bob begin by using local unitaries to rotate the Schmidt
basis of each $\ket{\psi_i}$ into a standard basis.  This leaves them
with the state 
$$
\sum_{j_1=1}^d \sqrt{p^1_{j_1}}\ket{j_1}^A\ket{j_1}^B 
\ot \cdots\ot
\sum_{j_n=1}^d \sqrt{p^n_{j_n}}\ket{j_n}^A\ket{j_n}^B 
$$
where $(p^i_1,\ldots,p^i_d)$ are the Schmidt coefficients of
$\ket{\psi_i}$ (possibly not all nonzero).

Next they each project onto the subspace where the Schmidt
coefficients are in the range $\exp(-E\pm n\delta/2)$; that is spanned by
$\ket{j_1,\ldots,j_n} \ot \ket{j_1,\ldots,j_n}$  for those values of
$j_1,\ldots,j_n$ satisfying $|\sum_i\log p^i_{j_i} + E| \leq
n\delta/2$.  We will later argue that this projection almost always
succeeds, and thus causes very little disturbance.

Now they divide the interval $[\exp(-E-n\delta/2),\exp(-E+n\delta/2)]$
into $m$ bins with geometrically spaced boundaries, for $m$ a
parameter we will pick later.  That is, for
$k=1,\ldots,m$, bin $k$ is the interval
$$\l[\exp\!\l(\!-E\!-\!\frac{n\delta}{2}\!+\!\frac{j-1}{m}n\delta\!\r),
\exp\!\l(\!-E\!-\!\frac{n\delta}{2}\!+\!\frac{j}{m}n\delta\!\r)\!\r].$$
Still without using any communication, Alice and Bob each perform a
projective measurement onto the different bins.   By this we mean that
the measurement operators are projectors onto subspaces spanned by
strings $\ket{j_1,\ldots,j_n}$ whose Schmidt
coefficients fall entirely into one of the above intervals.
We claim that (a)
that with high probability they will find a bin that contains many
eigenstates, and (b) the resulting state will have high fidelity with
a maximally entangled state.  It remains only to quantify the various
errors and inefficiencies we have encountered along the way.

Start with the last step.  The probability that the bin measurement
yields a bin with weight $\leq \eps$ (for $\eps$ a parameter we will
set later) is $\leq m\eps$.  Each Schmidt coefficient that we have
kept is $\leq \exp(-E+n\delta/2)$, and thus any bin with weight $\geq
\eps$ must contain $\geq \eps\exp(E-n\delta/2)$ eigenvalues.  Choosing
$\eps=2^{-n\delta/2}$, we obtain a state that approximates
$\geq \log 1/\eps + E-n\delta/2 = E-n\delta$ EPR pairs.  To assess the
fidelity of this approximation, note that all of the Schmidt
coefficients of the projected and rescaled state are in a band between
$\lambda$ and $\lambda\exp(-n\delta/m)\geq\lambda(1-n\delta(\ln 2)/m)$, for some
normalization factor $\lambda$.  Thus this state
has fidelity $1-O(n\delta/m)$ with a maximally entangled state.  Now choose
$m=2^{n\delta/4}$, so that both the $m\eps$ failure probability and
the $O(n\delta/m)$ error are $\leq\exp(-n\delta/5)$.

All that remains for the error analysis is to assess the damage from
projecting onto the 
Schmidt coefficients in the interval $-E\pm n\delta/2$.  The key tool
here is the Chernoff bound\cite{Chernoff52}, which states that if
$X_1,\ldots,X_n$ are independent (not necessarily independent) random
variables satisfying $0\leq 
X_i \leq \gamma$ then $X:=\sum_{i=1}^nX_i$ satisfies
\be P\l(|X-\bbE X| \geq n\frac{\delta}{2}\r) \leq 2\exp\l(-\frac{n\delta^2}
{\gamma^22\ln 2}\r).\label{eq:chernoff}\ee
We would like to apply this with $X_i$ defined by $P(X_i=\log 1/p^i_j)
= p^i_j$, so that $X$ is likely to be close to $\bbE X=E$.
Unfortunately, $p^i_j$ can be arbitrarily close to zero, and thus $\log
1/p^i_j$ can be arbitrarily large, so we cannot immediately establish any upper
bound on $\gamma$\footnote{The Chebyshev bound would avoid these
  difficulties, but at the cost of losing the exponential bounds on
  error probability.  Nevertheless, for small values of $n$, it may be
preferable.  Here we can use the fact that $\Var(X)\leq n\log^2 d$ to
find that $P(|X-\bbE X| \geq n\delta/2) \leq 4(\log
d)^2/n\delta^2$.  Other than the revised error bound, the rest of the
proof would be the same.}.
To do so, we will discard the Schmidt coefficients
that are smaller than $2^{-\gamma}$, which automatically means that
$0\leq X_i \leq \gamma$.  This causes $\leq\! d2^{-\gamma}$ damage for
each $\ket{\psi_i}$, or $\leq\! nd2^{-\gamma}$ overall.  
Combining with \eq{chernoff}, we find the total error is
$nd2^{-\gamma}+e^{\frac{-n\delta^2}{2\gamma^2}}$.  Finally we set
$\gamma = (n\delta^2)^{1/3}$ to obtain an overall error of
$O(\exp(-(n\delta^2)^{1/3}/2\ln 2))$, which dominates the 
$\exp(-n\delta/5)$ 
error from the first part.

Now we explain how the protocol can be made semi-clean.  The ``which
bin'' measurement should instead be performed coherently, with the
entanglement extraction proceeding conditioned on the quantum register
storing the superposition of measurement outcomes.  Since the
different outcomes remain locally orthogonal, the overall Schmidt rank
is equal to the sum over $k$ of the rank of the state conditioned on
obtaining bin $k$.
After 
the initial projection onto the typical subspace succeeds, each
Schmidt coefficient is $\geq \exp(-E-n\delta/2)$, and so each bin
has rank $\leq \exp(E+n\delta/2)$.
Since we are
extracting $E-n\delta$ ebits from each bin, the rank of the residual
state, conditioned on $k$, should be $\leq
\exp\l(\frac{3}{2}n\delta\r)$.
Once we sum over $m=2^{n\delta/4}$ bins, we have an overall Schmidt
rank of $\leq \exp(\frac{7}{4}n\delta)\leq \exp(2n\delta)$.
\endproof

\myproof{of coding theorem for \thm{C-E-toff}}
Suppose there exists an ensemble $\cE^{XAA'BB'} = \sum_x p_x\proj{x}^X
\ot \psi_x^{AA'BB'}$ such that
\be\begin{split} I(X ; BB')_{U(\cE)} - I(X ; BB')_{\cE} &= C
\\\text{ and }
H(BB'|X)_{U(\cE)} - H(BB'|X)_{\cE} &= E,\end{split}\ee
The idea is to use HSW coding\cites{Holevo98,SW97} for
$U(\cE)^{BB'}$.  For a string $\vec{x}=(x_1,\ldots,x_n)$, let
$\ket{\psi_\vx} := \ket{\psi_{x_1}} \ot \ldots \ket{\psi_{x_n}}$.
The HSW theorem states that for $\eps,\delta>0$ and for $n$
sufficiently large, choosing $N:=\exp(n(I(X;BB')_{U(\cE)}-\delta))$ 
random codewords $U^{\ot n}(\psi_{\vx_1})^{BB'},\ldots, U^{\ot
  n}(\psi_{\vx_N})^{BB'}$ according to the distribution
$\vec{p}(\vx):=p_{x_1}p_{x_2}\ldots p_{x_n}$ will result in a code
with average error $\leq \eps$.

On the other hand, the operator Chernoff bound\cite{Winter99} states
that a collection of $M:=\exp(n(I(X;BB')_\cE + \delta))$ random
codewords $\psi_{\vx_1},\ldots,\psi_{\vx_M}$ will have average state
on Bob's side quite close to their expectation
\be\theta^{BB'} := \bbE_\vx \psi_\vx^{BB'} = 
\l(\sum_x p_x \psi_x^{BB'}\r)^{\ot n}.\ee
Choose an arbitrary purification $\ket{\theta}^{ABB'}$.
Also let $N=LM$, so $L=\exp(n(I(X;BB')_{U(\cE)} - I(X;BB')_\cE -
2\delta)) = \exp(n(C-2\delta))$.  Our strategy for the rest of the
proof is for Alice and 
Bob to start with a state where Bob's part always looks like
$\theta^{BB'}$, but Alice can reliably send one of $L$ different
messages by performing a local unitary and then applying $U$ to the
joint state.

With this in mind, we now rephrase the random codes described above.
Draw $\{\vx_{i,j}\}_{i\in[L],j\in[M]}$ from the distribution $\vp$,
and let $e_{i,j}$ be the probability of error when Bob attempts to
decode $U^{\ot n}\ket{\psi_{\vx_{i,j}}}$.  The HSW theorem states that
with high probability the average error is low, i.e.
\be \frac{1}{LM}\sum_{i=1}^L\sum_{j=1}^M e_{i,j} \leq \eps.
\label{eq:hsw-good}\ee
On the other hand, the operator Chernoff bound states that with high
probability
\be \frac{1}{M}\sum_{j=1}^M \psi_{\vx_{i,j}}^{BB'} 
\approx_{\eps^2/2} \theta^{BB'}\label{eq:chernoff-good}\ee
for all $i\in[L]$ (the reason to demand error $\eps^2/2$ will later be
apparent).  Using the union bound (see e.g. the proof of
Theorem~1 of \cite{Devetak03} for detailed calculations), one can show
that in fact with high probability both \eqs{hsw-good}{chernoff-good}
hold simultaneously, and in particular that there exists a set of
$\{\vx_{i,j}\}$ for which this is true.  Fix this set for the rest of
the proof.

 For each $i$, let $e_i := \sum_j e_{i,j} / M$ and define the set of good
codewords to be $G = \{i :
e_i \leq 2\eps\}$.  By Markov's inequality, $|G|\geq
L/2$.\footnote{Note that unlike in standard HSW coding, we cannot
simply throw out 
the worst half of all codewords, since then the $\vx_{i,j}$ would nol
onger be independent and \eq{chernoff-good} would no longer
necessarily hold.}
The communication protocol proceeds as follows:
\benum \item
Alice and Bob start with the state $\ket{\theta}^{ABB'}$.
\item
To send the message $i\in G$, Alice will perform a local unitary
operation so 
that the overall state is within $\eps$ of 
$$\frac{1}{\sqrt{M}} \ket{i}^A \sum_{j=1}^M \ket{j}^A 
\ket{\psi_{\vx_{i,j}}}^{AA'BB'}.$$
This is possible because of \eq{chernoff-good}, Uhlmann's
theorem\cite{NC00}, and the fact that two mixed states with trace
distance $\leq\eps^2/2$ have purifications with trace distance $\leq
\eps$ \cite{DHW05}*{Lemma 2.2}.

\item Apply $(U^{AB})^{\ot n}$ so that the two parties share a state
within $\eps$ of 
$$\frac{1}{\sqrt{M}} \ket{i}^A \sum_{j=1}^M \ket{j}^A 
(U^{AB} \ot I)^{\ot n}\ket{\psi_{\vx_{i,j}}}^{AA'BB'}.$$

\item Bob decodes coherently, to obtain a state within $\eps + e_i
\leq 3\eps$ of
$$\frac{1}{\sqrt{M}} \ket{ii}^{AB}
\sum_{j=1}^M \ket{jj}^{AB}
(U^{AB} \ot I)^{\ot n}\ket{\psi_{\vx_{i,j}}}^{AA'BB'}.
$$

\item Conditioned on $i,j$, Alice and Bob concentrate $\approx
nH(BB'|X)_{U(\cE)}$ ebits from 
$(U^{AB} \ot I)^{\ot n}\ket{\psi_{\vx_{i,j}}}^{AA'BB'}$.  
Since the dimension of the states is fixed and $n$ can be made
arbitrarily large, the entanglement concentration technique of
\cite{BBPS96} will suffice.  
Moreover, entanglement concentration can be performed cleanly, so a
sublinear 
amount of additional communication will leave them with the state
$M^{-1/2}\sum_{j=1}^M \ket{j}^A \ket{j}^B$, which is of course
equivalent to $n(I(X;BB')_\cE+\delta)$ ebits. 
\eenum

Alice and Bob have used $U$ $n + o(n)$ times and sent $\log L =
n(C-2\delta)$ 
cobits.  They started with the state $\ket{\theta}^{ABB'}$, which can
be prepared with entanglement dilution using $nH(BB')_\cE + o(n)$
ebits and $o(n)$ cbits\cite{LP99}, and end with $n(I(X;BB')_\cE +
H(BB'|X)_{U(\cE)}) \pm o(n)$ ebits, for a net change of
$n(H(BB'|X)_{U(\cE)} - H(BB'|X)_{\cE}) \pm o(n) = n(E \pm o(1))$, as
desired.
\endproof

Note that unlike the proof of \thm{std-form} (or indeed the proof in
\cite{Har03} of the present result), no double-blocking is necessary
here, except perhaps to deal with the sublinear communication used for
entanglement dilution and to erase the ancilla states left by
entanglement concentration.


\DefinePublisher{ams}{AMS}{American Mathematical Society}{Providence}

\DefineJournal{ieeeit}{IEEE Trans. Inf. Theory}
{0018-9448}{IEEE Transactions on Information Theory}

\DefineJournal{pra}{Phys. Rev. A}
{0000-0000}{Physical Review A}

\DefineJournal{prl}{Phys. Rev. Lett.}
{0000-0000}{Physical Review  Letters}

\DefineJournal{qic}{Quantum Info. Comp.}
{0000-0000}{Quantum Information and Computation}

\DefineJournal{cmp}{Comm. Math. Phys.}
{0000-0000}{Communications in Mathematical Physics}

\DefineJournal{jmp}{J. Math. Phys.}
{0000-0000}{Journal of Mathematical Physics}

\end{document}